\documentclass[sn-basic, Numbered]{sn-jnl}

\usepackage{graphicx}%
\usepackage{multirow}%
\usepackage{amsmath,amssymb,amsfonts}%
\usepackage{amsthm}%
\usepackage{mathrsfs}%
\usepackage{mathtools}
\usepackage[title]{appendix}%
\usepackage{xcolor}%
\usepackage{textcomp}%
\usepackage{manyfoot}%
\usepackage{booktabs}%
\usepackage{algorithm}%
\usepackage{algorithmicx}%
\usepackage{algpseudocode}%
\usepackage{listings}%

\usepackage{float}
\usepackage{subcaption} 
\usepackage{tikz}

\tikzstyle myBG=[line width=3pt,opacity=1.0]

\newcommand{\R}{\mathbb{R}}

\newcommand{\Z}{\mathbb{Z}}

\newcommand{\ux}{\mathbf{x}}
\DeclareMathOperator*{\argmin}{arg\,min}
\newcommand{\dx}{\,\text{d}\ux}

\newcommand{\ub}{\mathbf{u}}
\newcommand{\vb}{\mathbf{v}}
\newcommand{\pb}{\mathbf{p}}

\newcommand{\prx}[1]{\text{prox}_{\tau F}(#1)}
\DeclareMathAlphabet{\pazocal}{OMS}{zplm}{m}{n}

\newcommand{\zpftext}[1]{\hspace{0.5cm}\text{#1}\hspace{0.5cm}}

\newcommand{\J}{\pazocal{J}}
\newcommand{\LL}{L}

\newcommand{\dvg}{\,\textnormal{div}\,}

\newcommand{\prox}{\text{prox}}

\newcommand{\evn}{\text{even}}
\newcommand{\odd}{\text{odd}}

\newcommand*{\inlg}[1]{%
    \raisebox{-.2\baselineskip}{%
        \includegraphics[
        height=0.8\baselineskip,
        width=0.8\baselineskip,
        keepaspectratio,
        ]{#1}%
    }%
}

\theoremstyle{thmstyleone}%
\theoremstyle{thmstyletwo}%

\theoremstyle{thmstylethree}%

\begin{document}

\title[MorphFlow]{MorphFlow: Estimating Motion in In Situ Tests of Concrete}


\author*[1]{\fnm{Tessa} \sur{Nogatz}}\email{tessa.nogatz@rptu.de}

\author[1]{\fnm{Claudia} \sur{Redenbach}}\email{claudia.redenbach@rptu.de}
\equalcont{These authors contributed equally to this work.}

\author[2]{\fnm{Katja} \sur{Schladitz}}\email{katja.schladitz@itwm.fraunhofer.de}
\equalcont{These authors contributed equally to this work.}

\affil[1]{\orgdiv{Mathematics Department}, \orgname{RPTU Kaiserslautern-Landau}, \orgaddress{\street{Gottlieb-Daimler-Straße 48}, \city{Kaiserslautern}, \postcode{67663}, \country{Germany}}}

\affil[2]{\orgdiv{Image Processing Department}, \orgname{Fraunhofer ITWM}, \orgaddress{\street{Fraunhofer-Platz 1}, \postcode{67663}, \city{Kaiserslautern}, \country{Germany}}}


\abstract{We present a novel algorithm explicitly tailored to estimate motion from time series of 3D images of concrete. Such volumetric images are usually acquired by Computed Tomography and can contain for example in situ tests, or more complex procedures like self-healing. Our algorithm is specifically designed to tackle the challenge of large scale in situ investigations of concrete. That means it cannot only cope with big images, but also with discontinuous displacement fields that often occur in in situ tests of concrete.
We show the superior performance of our algorithm, especially regarding plausibility and time efficient processing. Core of the algorithm is a novel multiscale representation based on morphological wavelets. We use two examples for validation: A classical in situ test on refractory concrete and and a three point bending test on normal concrete. We show that for both applications structural changes like crack initiation can be already found at low scales -- a central achievement of our algorithm.}

\keywords{X-Ray Computed Tomography, In situ, DVC, Refractorary Concrete, Glass Fiber Reinforced Concrete}

\maketitle

\section{Introduction}
\label{Intro}
Concrete is for sure among the most important materials in construction. Especially its use in critical infrastructure like bridges requires a thorough assessment of the material's behavior. A very common way to do so is by situ materials tests. Such tests couple any kind of materials test with an imaging modality. Using computed tomography (CT) as modality is of special interest as it allows to get a complete, three-dimensional digitized version of the whole concrete sample. This allows to assess the inside of the specimen and -- when combined with a materials test -- also its response to load. In situ materials tests follow a specific procedure: A sample is scanned without any load in a CT device. Then, ideally without removing the sample from the CT, a limited amount of load is applied. The load is kept constant and after a suitable time period of relaxation, the sample is scanned again. This is repeated until the desired material failure happens. This way, a time series of three-dimensional images is generated.

When applied to concrete, such in situ tests have already proven to increase the understanding of fracture behavior. The authors of \cite{Tsitova_2022} for example used in situ tests to validate numerical predictions of cracking areas. However, concrete in situ tests are yet particularly difficult to evaluate. In our work, we focus on tackling two specific difficulties. First, due to its brittle nature, concrete exhibits discontinuous behavior. Its response to load is mainly fracture or cracking  -- a pattern that many state-of-the-art motion estimation algorithms in materials science fail to display. Second, there is a rising demand in efficient processing of large scale data. Tomographs like the tomography portal Gulliver\footnote{https://bauing.rptu.de/ags/massivbau/forschung/dfg-grossgeraeteinitiative} produce images of a size of  several hundreds of gigabytes or even some terrabytes. Efficient processing is crucial in these applications. Unfortunately, crack segmentation in concrete is a very costly task, even when Deep Learning methods are applied \cite{Barisin_2022}. One therefore is in need for a preprocessing technique, that indicated areas of potential cracks and \emph{guides} the costly methods to positions where cracks can potentially be found. Therefore, a special focus shall be put on working with compressed and therefore much smaller data.

State-of-the-art algorithms for computing the motion between each element of the time series in general go under the name \emph{Digital Volume Correlation (DVC)}, which was introduced first by Bay et al. in 1999 \cite{bay}. We recently showed that voxel-accurate displacement estimation based on Optical Flow (OF) formulations outperforms modern and classical DVC methods \cite{strain}. As this is mainly due to their capability of capturing discontinuous behavior such as cracking or fracture, they are also especially suitable to estimate displacement in in situ tests of concrete. OF methods naturally incorporate a multiscale approach, that means the problem of motion estimation is solved on a coarse scale and the solution is then used as initial point for a solution on a finer scale. We make use of this feature and exchange the classical Gaussian coarse-to-fine scheme by an image pyramid based on morphological wavelets \cite{Heijmans_2000}. As wavelets are also applied for compressing data, using these within our algorithm satisfies our second need -- the efficient processing of large image data.

However, the algorithm that we used in our previous work \cite{strain} contains a highly involved solution scheme based on successive overrelaxation. The incorporation of morphological wavelets is not straight-forward, we therefore decided to choose a variant of OF that is easier to implement, namely the TV-$L_1$ algorithm by Zach et al. \cite{Zach_2007}.

Besides the utilization of compressed data, there is a further reason to exchange the common Gaussian pyramids in OF by morphological wavelets. Traditionally, Gaussian pyramids have been used because one can show their optimality with respect to images of natural scenes, see for example \cite{Lindeberg_1994}.  
Gaussian smoothing and subsampling reduce small scale details. Combined with motion estimation, this follows the assumption that motion of large details in the image dominates also the motion of small details. In classical digital images, it is not too harmful to blur and compute motion based on large objects first. In an in situ test on concrete, however, the dominant feature that indicates the motion is the crack. And this feature \emph{will} be blurred by Gaussian smoothing.

Morphological wavelets overcome this problem. The wavelet that we use is a rather uncommon representative of the so-called second generation wavelets. Wavelets in their proper sense are actually not really suitable for replacing Gaussian pyramids. Classical wavelets are supposed to preserve as much detail as possible from one scale to the next. In OF, however, it is required that part of the detail is lost in between scales to prevent algorithms from being trapped in local minima. We show that morphological wavelets exhibit this behavior much better than conventional wavelets of first and second generation.

\begin{figure}
		\subfloat[Unloaded refractory concrete]{   
			\includegraphics[width=0.445\textwidth]{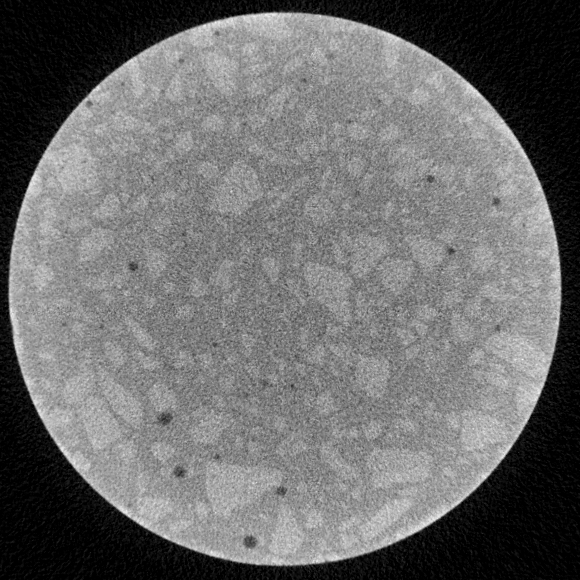}
			\label{KDF0sl}
		}\hspace{0.5cm}
		\subfloat[Loaded refractory concrete]{
			\includegraphics[width=0.45\textwidth]{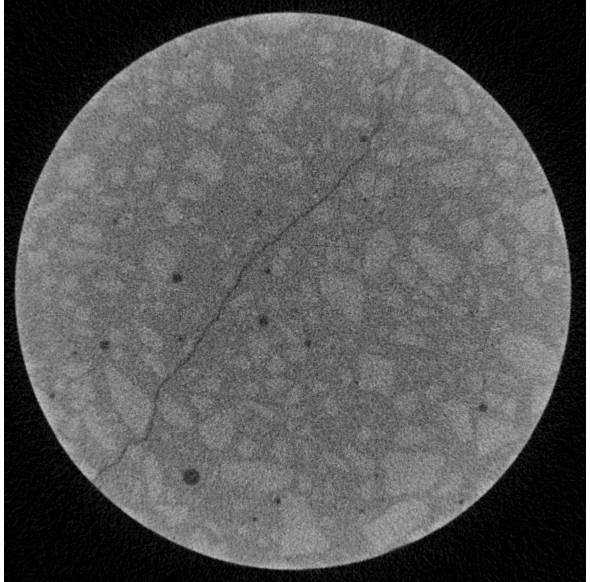}
			\label{KDF1sl}
		}
			\\
			\subfloat[Unloaded reinforced concrete]{
				\includegraphics[width=0.45\textwidth]{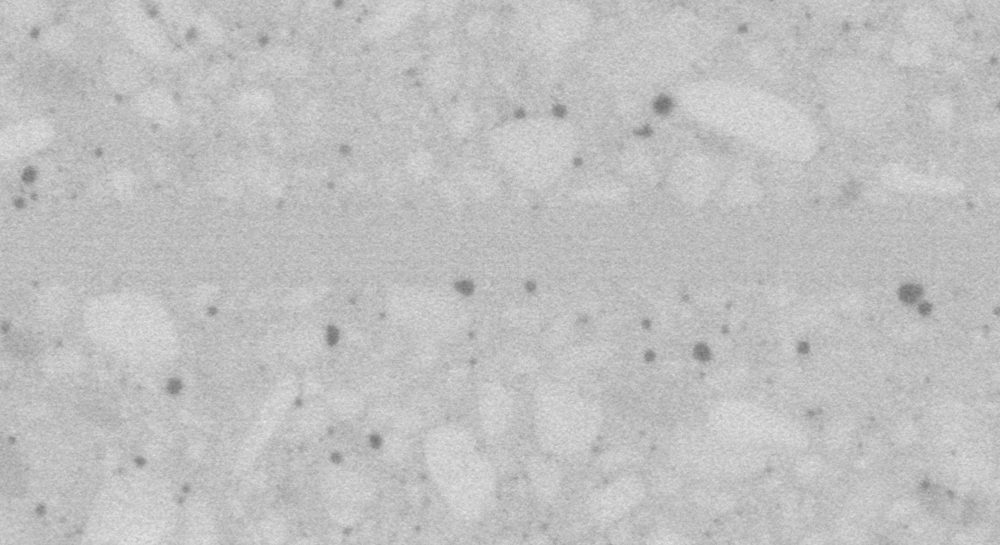}
				\label{gfk0sl}
			}\hspace{0.5cm}
		  \subfloat[Loaded reinforced concrete]{
			\includegraphics[width=0.45\textwidth]{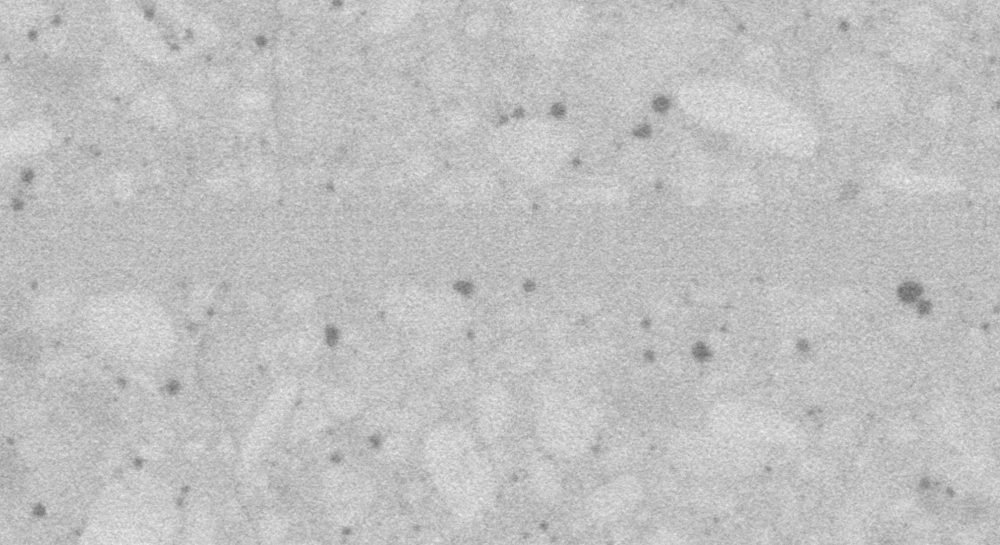}
			\label{gfk2sl}
        }
		
	\caption[Sliceviews of used datasets]{Sliceviews of used data sets. (\protect\subref{KDF0sl}) and (\protect\subref{KDF1sl}) show two slices of the in situ test on refractory concrete. Experiment and imaging: TUBAF Freiberg. The images have a size of $580\times 580\times 638$ voxels at a voxel edge length of 50 $\mu$m.  (\protect\subref{gfk0sl}) shows the sliceview of the reinforced concrete before loading. Although \protect\subref{gfk2sl}) already depicts a loading stage, the cracks are difficult to spot by human eyesight. The image has a size of $550\times 835 \times 1000$ voxels with voxel edge length of $125\mu$. CT scanning: Fraunhofer EZRT, Fürth. Experiment: Civil Engineering, RPTU Kaiserslautern-Landau, Kaiserslautern.}
	\label{datasl}
\end{figure}

We show the power of our algorithm with two in situ tests of concrete. The first one is a classical in situ test of refractory concrete. The second one is a more involved three point bending test on reinforced concrete. Reinforcement is provided by a glass fiber rebar. Sliceviews of the images of both samples are shown in Figure~\ref{datasl}.

The remainder of this manuscript is organized as follows. In Section~\ref{RelWork} we give a short review on in situ tests on concrete and wavelet based motion estimation. In Section~\ref{OF} we present the plain motion estimation algorithm, which is a 3D extension of a classical algorithm from digital image processing, without any comments on the multiresolution decomposition. This is presented in detail in Section~\ref{MorphWave}, where morphological wavelets are defined and their use in the implementation of OF is outlined. Section~\ref{Datasets} briefly introduces data sets and methodology used for evaluation. The results of the evaluation are presented in Section~\ref{Eval}. Section~\ref{Conclusion} contains concluding remarks.

\section{Related Work}
\label{RelWork}
\subsection{In Situ Investigations on Concrete}
In situ investigations of concrete require a delicate planning of the testing procedure. Concrete is a brittle material meaning that initiation of damage such as cracks happens abruptly after a short elastic phase. Due to this behaviour, it is very difficult to fix the loading stages beforehand. Therefore, loading is often increased stepwise, and an external pressure management is applied to decide whether the concrete failed or not. Unfortunately, microcracking does not influence the mechanical properties notably. Therefore, when failure is detected by external monitoring, the concrete usually already exhibits severe, large cracks.

Nevertheless, various in situ investigations of concrete can be found. They are performed either in laboratory CT devices or at synchrontron beamlines. Many of them apply some kind of motion estimation, mainly DVC,  to assess the loading induced motion and eventually the crack. DVC does not compute displacement per voxel but per subvolume. \emph{Local} approaches fix independent subvolumes and calculate an average displacement for each subvolume independently from the surrounding ones, \emph{global} methods use finite element grids as underlying subvolumes and calculate nodal displacements.

\subsubsection*{Analysis by Local DVC}
Contributions that specifically use local DVC methods for concrete data are for example \cite{Chateau_2018, lorenzoni_2020, mao_2019, Trtik_2007, Yang_2013}. The main problem that arises when using local DVC based methods is that they often fail to display the crack in the displacement field. This is especially visible in the manuscripts \cite{lorenzoni_2020} and \cite{mao_2019}, where the results rather show smooth fields instead of jumps, as one would expect if a crack forms. Even further, Chateau et al. \cite{Chateau_2018} use the poor performance of DVC at cracks to segment displacement based on errors in residuals.

Local DVC was also applied to concrete for estimating motion that was not induced by load but by environmental influences. Chavez et al. \cite{chavez2020real} investigated cement carbonation while Jiang et al. \cite{jiang2020simulation} studied corrosion induced cracking. 

\subsubsection*{Analysis by Global DVC}
Global methods are usually not available open-source. Therefore, one of the few works combining in situ tests on concrete and global DVC is quite a recent one by Tsitova et al. \cite{Tsitova_2022}. It shows many of the struggles that come with using global DVC for brittle materials such as concrete. As global DVC produces rather smooth displacement fields, it tends to smear out the strain extrema that one would expect at cracks. The authors of \cite{Tsitova_2022} had to apply quite heavy post-processing to deduce information in cracked areas nevertheless. They extracted the cracks by erroneous residuals and incorporated this information in \emph{another} iteration of motion estimation. 

\subsubsection*{Analysis without Motion Estimation}
Unfortunately, there are also several contributions that do not fully exploit the underlying data, as they only visualize the data and do not apply any motion estimation algorithm. Among them are \cite{Huang_2016, Landis_2009, Yang_2020}

\subsection{Wavelets in Multiresolution Schemes}

\subsubsection*{General Images}

In motion estimation and specifically OF wavelets have so far mainly been used to decompose the displacement field rather than the image itself.
Instead of calculating a displacement vector for every voxel, one formulates the field in terms of a wavelet basis. Conceptwise, the approach is comparable to using an FE grid as in DVC or even classical FE simulations. However, using wavelet bases allows for much more freedom from a mathematical point of view in terms of the underlying nature of the displacement fields. That is why these approaches perform reasonably well compared to classical OF approaches \cite{Wu_1998, Derian_2013}.

When it comes to the multiresolution decomposition, the use of wavelets is quite rare, as wavelets are not supposed to smooth from one scale to the next. Nevertheless, one can find some very advanced applications mainly in the area of ultrasound images. Saputro et al. \cite{Saputro_2010} propose to use a wavelet based multiresolution approach to support OF computation for myocardial motion. Gastounioti et al. \cite{Gastounioti_2011} aim for arterial wall displacements. Both applications deal with 2D images. To the best of our knowledge, no 3D applications have been reported so far.

\subsubsection*{CT Images}

Wavelets have sucessfully been applied to concrete in several applications. Similar to classical image processing, compression \cite{stock2020edge} and denoising \cite{borsdorf2008wavelet} approaches by wavelets can be found. Wavelets have also been used for segmenting \cite{padma2014segmentation} and classifying CT data \cite{hosntalab2010classification}. Further, one can find them also in artifact reduction \cite{mehranian2013x}.

\section{Optical Flow for Volume Images}
\label{OF}

Our method employs concepts from various areas that are combined to solve a very special class of problems. The core of our algorithm is an extension of a variant of total variation regularized OF to 3D. Contrary to our previous publication \cite{strain},  we derive our motion estimation from a different variant of optical flow, namely that of Zach et al. \cite{Zach_2007}. Similar to 3DOF it has not been applied in materials science, but in medical imaging by Hermann et al. \cite{Hermann_2013}. Note that there are many ways to solve and formulate OF, as it is one of the core problems of Computer Vision. Methods do not only differ in the underlying equations but also in solution schemes. The method of Zach et al. requires a less advanced scheme than the one of Brox. As the the incorporation of our coarse-to-fine scheme is already quite demanding, we chose the former due to its easier solution scheme.

The general assumption of OF is that grey values in images are preserved under displacement. That means that for images $I_0, I_1 \colon \Omega \subset \R^3\rightarrow\R$ we have
\begin{equation}
	\label{OFIntro}
	I_0(\ux) = I_1(\ux+\ub(\ux))
\end{equation}

with a suitable displacement $\ub\colon \Omega\subset\R^3\rightarrow\R^3$. Reconstructing the displacement via the above equation is impossible. It is a so called ill-posed problem, where the degrees of freedom outnumber the known data points by far. In each voxel, we only have a scalar correspondence but want to derive a three-dimensional displacement vector. To overcome this problem, one first formulates an equivalent minimization problem and then adds an additional \emph{regularizing} term to stabilize the computations. In the case of TV-$\LL_1$ of Zach et al., regularization is by total variation. The problem then reads 

\begin{equation}
	\min_{\ub}\int_\Omega \lambda \vert I_0(\ux)-I_1(\ux+\ub)\vert+ \vert\nabla\ub\vert\dx.
\end{equation}

Combining the $\LL_1$ norm and total variation regularization to compute motion unites methods that have been proven to be favorable in image processing for quite some time. The use of $\LL_1$-norms is supported by the findings from compressed sensing (see for example \cite{kutyniok2014compressed}), which indicate that even for noisy measurements signals contain much more information than needed for reconstruction. Regularization by total variation removes spurious oscillations while maintaining sharp edges \cite{rudin1992nonlinear}. Note that this formulation nevertheless comes with obstacles.  
Due to the use of $\LL_1$-norms, the problem is neither linear nor convex nor differentiable. Several steps are required to transform the problem into a feasible formulation. The first step is to  perform a first order Taylor approximation, that is

\begin{equation}
	I_1(\ux+\ub) = I_1(\ux+\ub_0) + \nabla I_1^T(\ux+\ub_0)(\ub-\ub_0),
\end{equation}

to remove the nonlinearity introduced by $\ub$ being evaluated \emph{inside} the image.
Plugging this into the above functional, we get
\begin{equation}
	\min_{\ub} \int_\Omega\ \lambda \vert(\nabla I_1)^T\ub + I_1(\ux+\ub_0)- (\nabla I_1)^T\ub_0 - I_0(\ux) \vert + \vert\nabla \ub\vert\dx.
\end{equation}

The use of first order Taylor approximations to linearize the problem is one of the central reasons, why a coarse-to-fine scheme needs to be employed in almost all OF algorithms. Taylor expansions are good approximations only in a short radius around the point of interest. That means $\ub_0$ needs to be close to $\ub$, and \emph{close} in image processing means \emph{closer than one voxel}. For motion estimation on the original scale this sounds not very reasonable. But if the image is downscaled rigorously, the requirement of one voxel displacement can be easily met. If the coarse displacement is then used as starting point on the next finer scale, the assumption of small displacements is met again. 

Yet we remain with several nondifferentiable parts in our formulation, namely every time $\vert \cdot \vert$ is used.
Thus, \cite{Zach_2007}, Zach et al. propose to split the above equation into two parts: One dealing with motion estimation based on the image data, and another one with reconstructing a solution that satisfies the regularizer best. Thus, we introduce an auxiliary variable $\ub$ and solve

\begin{equation}
	\label{primalDualAux}
	\min_{\ub, \vb}\J_\theta(\ub, \vb) = \int_\Omega \vert\nabla \ub\vert + \frac{1}{2\theta}\|\ub-\vb\|_2^2 + \lambda \vert\rho(\vb)\vert\dx,
\end{equation}

where 
\begin{equation}
\label{rho}
\rho(\vb)=(\nabla I_1)^T\vb + I_0(\ux+\vb_0)- (\nabla I_1)^T\vb_0
\end{equation}
for some initial but known $\vb_0$. Equation~\eqref{primalDualAux} is then solved by alternating between the solution concerning the $\ub$ and the $\vb$ parts

\begin{align}
	\label{TVsub}\ub^{k+1} &= \min_{\ub} \int_\Omega \vert\nabla \ub\vert + \frac{1}{2\theta} \|\ub-\vb^k\|_2^2 \dx\\
	\label{imgsub}\vb^{k+1} &= \min_{\vb} \int_\Omega \|\ub^{k+1}-\vb\|_2^2 + \lambda \vert\rho(\vb)\vert\dx.
\end{align}

What seems rather odd in first place (solving two problems instead of one) eventually turns out to be a very elegant way to circumvent the problem of the non-differentiability of the $\LL_1$-norms.
Let us first fix the following notations:
\begin{equation}
    \label{notations}
	F(\cdot) = \vert\cdot\vert,  \hspace{0.7cm} K=\nabla, \zpftext{and} g(\ub) =  \frac{1}{2\theta}\|\ub-\vb\|_2^2.
\end{equation}
Let us then claim that $F$ is a convex, lower semicontinuous function, $K$ is a continuous linear operator and $G$ is continuous at $0$. For the choices in~\eqref{notations}, results can for example be found in the monograph \cite{bredies2011mathematische}. Then, for finite dimensional spaces $V$ and $Y$ we can deduce the saddle-point formulation

\begin{align}
	\min_{v\in V} F(K(v)) + g(v) & = \min_{v\in V}\sup_{y\in Y} \langle y, Kv\rangle - F^*(y) + g(v) \\
	&\label{maxinfpd} = \max_{y\in Y}\inf_{v\in V} \langle y, Kv\rangle - F^*(y) + g(v) 
\end{align} 

The choice for the finite dimensional space $V$ is a very obvious one: It is the one that contains our discrete displacement field, so when considering images of $N\times M\times L$ voxels, we have $V=\R^{N\times M\times L\times 3}$. The choice of $Y$ is a bit more delicate, but eventually one can see that $Y=\R^{N\times M\times L\times 3\times 3}$ is the correct choice. 

Note that we did not require $F$ to be differentiable. Therefore, necessary conditions for the existence of solutions to \eqref{maxinfpd} can only be posed in terms of the subdifferential. The subdifferential (if it exists) of a function $F$ at point $u$ is the set of all elements $u^*$ that satisfy
\begin{equation}
		F(v)\geq F(u) + <u^*, v-u>\;\;\forall v\in V.
	\end{equation}
Roughly speaking, the subdifferential can be treated like the gradient, but is now a set-valued quantity. Recall that a convex differentiable function has a minimum where its gradient equals $0$. This condition is necessary \emph{and} sufficient. In the case of operators that only have a subdifferential, the equality is now relaxed to $0$ being an element of the set-valued subdifferential of $F$. Nevertheless, this condition remains necessary and sufficient.

For our saddle-point formulation \eqref{maxinfpd} this results in
\begin{equation}
\label{subdiffcond}
\begin{split}
	0 &\in \partial g(u) + K^*y^*, \\
	0 &\in \partial F^*(y^*) - Ku,
\end{split}
\end{equation}
where we have a necessary condition for each of the unknowns $u$ and $y$. Note that the condition is also sufficient, as we deal with convex functions. 

Clearly, because we do not deal with differentiable functions, we cannot formulate a gradient descent method based on \eqref{subdiffcond}. What we can do nevertheless is to formulate proximal maps. As above, let $F$ again be a convex, lower semicontinuous function. Then, the proximal map $\prx{v}$ is defined by
\begin{equation}
	\label{critptprox}
	\prx{v}  = \argmin_u F(u) + \frac{\|u-v\|^2}{2\tau }.
\end{equation} 
Remarkably, this small modification in formulating the minimization problem now in fact allows to apply a gradient descent scheme. Based on \eqref{subdiffcond} we alternatingly compute

\begin{equation}
\begin{split}
	\pb^{k+1} &= \prox_{\sigma F^*} (\pb^k + \sigma K \bar \ub^k) \\
	\ub^{k+1} &= \prox_{\tau G}(\ub^k-\tau K^*\pb^{k+1}) \\
	\bar \ub^{k+1}&= \ub^{k+1}+ \theta(\ub^{k+1}-\ub^k)).
\end{split}
\end{equation}

Reformulation and calculation of the conjugate functions eventually results in

\begin{equation}
    \label{finalupdatep}
	\pb^{k+1} = \frac{\pb^k + \sigma \nabla \bar \ub^k}{\max(1, \left\vert\pb^k + \sigma K \bar \ub^k\right\vert)}
\end{equation}
and
\begin{equation}
    \label{finapupdatestepu}
	\ub^{k+1}  = \frac{\ub^k-\tau \dvg\pb^k+ \vb}{1+\tau\lambda},
\end{equation}
which means that the formerly non-differentiable problem can now be solved by iteration. 

It remains to solve \eqref{imgsub}. First of all, the problem is convex and therefore first order optimality is not only necessary but also sufficient to characterize a minimizer. Furthermore, the only non-differentiable part left is the absolute value $\vert \cdot \vert$ in the second summand. Recall, that the dependence  of $\vb$ in $\rho$ was only linear (see \eqref{rho}). That allows us to deduce a differentiable problem by case distinction. By considering the three cases $\rho(\vb)<0$, $\rho(\vb)>0$ and $\rho(\vb)=0$, we can derive from first order optimality that
\begin{equation}
    \label{finateupdatestepv}
	\vb^{k+1} = \ub^{k+1} +\begin{cases}\lambda\theta\nabla I_1\ &\text{if } \rho(\ub^{k+1}) < -\lambda \theta |\nabla I_1|^2 \\ 
		-\lambda\theta\nabla I_1\ &\text{if } \rho(\ub^{k+1}) > \lambda \theta |\nabla I_1|^2 \\
		-\rho(\ub^{k+1})\nabla I_1/|\nabla I_1|^2\ &\text{if } |\rho(\ub^{k+1})| \leq \lambda \theta |\nabla I_1|^2. 
	\end{cases}
\end{equation}

\section{Multiresolution Approximations}
\label{MorphWave}

In this section we will shortly introduce classical multiresolution analysis approaches. Let us first recapitulate the following general notion by Heijmans et al. \cite{Heijmans_2000}. A multiresolution analysis of a signal $f\in V_0$ consists of nested spaces $(V_j)_{j\in\mathbb{Z}}$, which satisfy the following properties
\begin{enumerate}
	\item $\cdots\subset V_2\subset V_1\subset V_0 \subset V_{-1}\subset V_{-2}\subset \cdots$, 
	\item $\bigcap_{j\in \mathbb{Z}} V_j = \{0\}$, $\overline{\bigcup_{j\in\mathbb{Z}}V_j} = \LL_2(\R)$, 
	\item $f\in V_j \leftrightarrow f(2^{-j}\cdot )\in V_{j+1}$, 
	\item $f\in V_j \rightarrow f(\cdot-2^jk) \in V_j$ for all $j,k\in\mathbb{Z}$.
\end{enumerate}
Increasing $j$ in this notation will decrease the information in the resulting space and therefore represents downscaling. 
To traverse from one scale to the next, multiresolution analyses usually come with \emph{analysis} and \emph{synthesis} operators. Analysis operators reduce the image content, synthesis operators restore the image content between levels with decreasing $j$.  
\subsection{Gaussian Pyramids}
\label{GaussianPyramids}

The aim of image pyramids is to encode the image information at different scales. Therefore, one is usually only interested in the analysis operator.   Following Adelson et al. \cite{Adelson_1984},  a level $0<\ell<N_L$ of a generic pyramid is computed via
\begin{equation}
    \label{pyramideq}
    G_\ell(i, j, k) = \sum_{m=-M}^{+M}\sum_{n=-N}^{+N}\sum_{l=-L}^{+L} w(m, n, l) G_{\ell-1}(2i+m, 2j+n, 2k+l),
\end{equation}
where $w$ is a suitable filter mask and the original image is set to be $G_0$. In the case of 3D Gaussian pyramids, $w$ is given by
\begin{equation}
    \label{kerneleq}
    w(m, n, l) = \frac{1}{(2\pi\sigma^2)^{3/2}}e^{\frac{-(m^2+n^2+l^2)}{2\sigma^2}}.
\end{equation}

Gaussian pyramids are therefore built by subsampling and convolutions with Gaussian kernels. That means that between each level of the pyramid structures of characteristic length $\sigma$ are blurred. Furthermore, due to the discrete nature of the application, \emph{filter masks} as in \eqref{kerneleq} are already assumed to have finite support by definition. The size of the support is usually chosen by the $2$-$\sigma$-rule, meaning that $M=2\sigma$ and $M=N=L$.

\subsection{Wavelets}
\label{Wavelets}
Wavelets, as the name indicates, are families of wave-like functions. They can be considered as an advancement of discrete Fourier transforms regarding their spatial localization. The Fourier transform is one of the oldest tools in image processing. It allows to decompose a signal or an image into its frequency parts. However, it is impossible to localize the frequency components. That means a small local change anywhere in the image changes the Fourier transform everywhere, so \emph{globally}. Furthermore, it is nearly impossible to distinguish the high-frequency components stemming from noise from those stemming from edges in the Fourier domain . Both aspects are not very desirable for a tool in image processing. Wavelets aim to overcome these two problems. They allow to localize the frequency components up to small variance, which is the best possible result due to Heisenbergs uncertainty principle. Furthermore, they have been used successfully in many imaging tasks, especially in denoising while preserving edges. 

Wavelets are formed via dyadic translates of one so called \emph{mother wavelet}, that is, all members of the family $\phi_{j,m}$ are computed via
\begin{equation}
    \label{motherwavelet}
	\phi_{j,m} = 2^{j/2}\phi(2^{j}-m).
\end{equation}

A~very old but well-known example are the \emph{Haar-wavelets}, whose mother wavelet is given by
\begin{equation}
	\label{HaarMother}
	\phi(x) = \begin{cases}
		1, & 0\leq x <\frac{1}{2},\\
		-1, & \frac{1}{2}\leq x < 1,\\
		0, &\text{otherwise.} 
	\end{cases}
\end{equation}

Clearly, the mother wavelet $\phi$ is just a step function on the interval $[0,1]$. The parameter $m$ now shifts the step function along the real axis. Positive $j$ now coarsen the intervals, whereas negative $j$ refine them. This way, the whole family of $\phi_{j,m}$ forms a basis, which we can use to approximate signals (or any function) $f$.  We have, up to a small error, 
\begin{equation}
	f = \sum_{j,m} \gamma_{j,m} \phi_{j,m},
\end{equation}
where $\gamma_{j,m}$ are the \emph{wavelet coefficients}.

\subsection{Lifted Wavelets}

Within the ``Wavelet-Boom'' between the late 1980s and early 2000s, many communities have proposed improvements and concepts in this area. A rather late one is the concept of \emph{Lifted Wavelets} that transfers the favorable properties of wavelets into a pure space domain decomposition. The Haar-wavelet in equation (\ref{HaarMother}) for example also has a pure space formulation: Instead of a continuous $f$, take its discretized version $f(k)$ sampled at a finite number of  points $k\in\Z$. (That is of course not a restriction in any real-world application.) The approximation $f_{1}$ of $f$ in the next coarser space is then

\begin{equation*}
	f_{1}(k) = \frac{f(2k)+f(2k+1)}{2},
\end{equation*}
and the wavelet coefficient is given by $\gamma_{1,k} = f(2k+1)-f(2k)$.

The idea of such lifted wavelets is that correlation of structures is already apparent in the space domain and a transfer to the frequency domain might therefore not be necessary. The scheme is best described by considering an abstract example. Once more consider a signal~$f$ which we know at a fixed sampling distance, e.g., $d=1$. This ``initial'' or ``original'' scale will be denoted by $\lambda_{0,k} = f(k)$, $k\in\Z$. An approximation in the next coarser space can now be to just consider the even samples, that it
\begin{equation}
	\lambda_{1,k} \coloneqq \lambda_{0,2k} \zpftext{for} k \in \Z.
\end{equation}
Ideally, the loss of information in such a downscaling step should be small. That means, that the difference between $\lambda_{1,k}$ and $\lambda_{0,k}$ is kept low. The coefficients, that will encode the difference, will be denoted by $\{\gamma_{1, k}\}$. 

In fact, these are, as in the classical approach to wavelets, the \emph{wavelet coefficients}. 

Now, a very obvious choice is to just put the odd coefficients as wavelet coefficients, that is $\gamma_{1, k} = \lambda_{0,2k+1}$, $k\in\Z$. This is also called the \emph{lazy wavelet}. However, this will not fulfill our requirement to produce small wavelet coefficients, even worse: the signal will be roughly the same as the approximation. Therefore, we need a more advanced scheme. 

We for now persist with the separation of even and odd samples. But now we \emph{predict} the odd samples $\{\lambda_{0,2k+1}\}$ based on $\{\lambda_{1,k}\}$ by computing the average of two (even) neighbors. The wavelet coefficient then reads

\begin{equation*}
	\gamma_{1, k} \coloneqq \lambda_{0,2k+1}- \frac{1}{2}\left(\lambda_{1,k}+ \lambda_{1,k+1}\right).
\end{equation*}

If the signal we encode contains correlated structures (which it does, as it will be an image), these coefficients are small. 
Nevertheless, there is still an issue with the $\lambda_{j,k}$'s. If we proceed by calculating them with the lazy wavelet up to the coarsest level, only the first and the last coefficient of the original signal will be left. However, it would be more intuitive, if the last coefficient for example resembled the average. This also makes the scheme less prone to aliasing. It can be achieved by requiring the average to be the same over all levels, i.e.

\begin{equation}
	\sum_k \lambda_{1,k} = \frac{1}{2} \sum_k \lambda_{0,k}.
\end{equation}
Clearly, the way to do so is to calculate averages: 

We use the neighboring \emph{wavelet} coefficients and \emph{lift} (or in other words update) the $\lambda_{1,k}$:
\begin{equation*}
	\lambda_{1,k} = \lambda_{1,k} + \frac{1}{4}\left(\gamma_{1, k-1} + \gamma_{1, k}\right).
\end{equation*}
Let us now formalize this procedure. The \emph{lifting scheme} consists of the following steps.
\begin{enumerate}
	\item Split the signal into subbands by an invertible transformation $\Sigma$. For the sake of simplicity, we use $2$ subbands. We have 
	\begin{equation*}
		(\evn_{1,k}, \odd_{1, k}) = \Sigma(\lambda_{0,k}).
	\end{equation*}
	\item Predict the wavelet coefficient (also called \emph{detail signal}) by $\odd_{1,k}$. We have
	\begin{equation*}
		\gamma_{1, k} = \odd_{1, k} - P(\evn_{1,k}),
	\end{equation*}
	where $P$ is an (invertible) prediction operator.
	\item Update the signal approximation $\lambda_{1,k}$ by the previously predicted wavelet coefficients
	\begin{equation*}
		\lambda_{1,k} = \evn_{1,k} + U(\gamma_{1, k}),
	\end{equation*}
	with update operator $U$.
\end{enumerate}

\subsection{Min/Max Lifting on the Quincunx Grid}

\begin{figure}
			\subfloat[3D Checkerboard]{  
            \label{checkerboard}
			\includegraphics[width=0.45\textwidth]{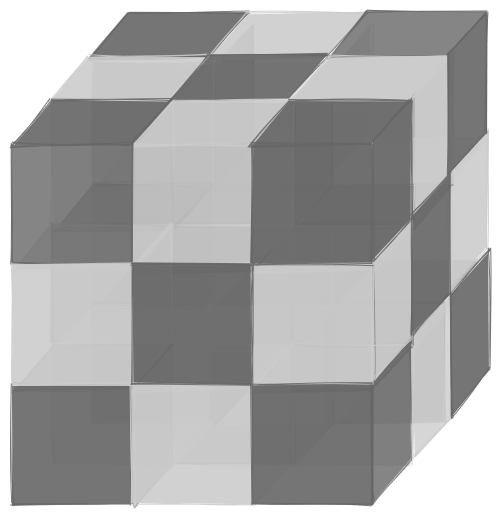}	
			}
			\hfill
			\subfloat[Quincunx Relation]{
            \label{check_qc}
            \includegraphics[width=0.45\textwidth]{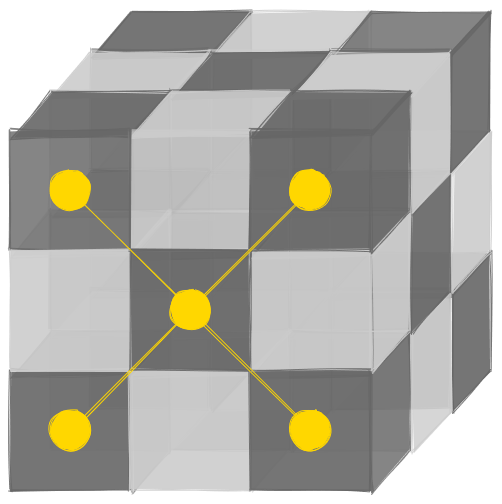}
			}
	\caption[3D Extension of even and odd]{3D Extension of separating a signal into even and odd resulting in a classical checkerboard pattern. Even and odd is now resembled by black and white. Neighboring relations among each label are now given by quincunxes, depicted in green in Figure~\protect\subref{check_qc}.}
	\label{Quincunx}
\end{figure}

The starting point for lifted wavelets for 1D signals in the last section was the Lazy wavelet, which just splits a signal into two distinct groups. In higher dimensions, \emph{even} and \emph{odd} samples can be replaced by \emph{black} and \emph{white} pixels or voxels in a checkerboard pattern, see Figure~\ref{checkerboard}. In a wavelet context, this is also called the quincunx scheme. \emph{Quincunx}, stemming from the latin words \emph{quinque} and \emph{uncia} meaning five coin, basically indicates a five point stencil. It is displayed in green in Figure~\ref{check_qc} and specifies the neighboring relations among black pixels or voxels only. 

From now on, we denote by \emph{image grid} or \emph{Cartesian grid} a grid whose nodes are the pixel or voxel center points. 

Following Heijmans et al. \cite{Heijmans_2000}, we now denote by $S$ the set of all original points, by $Q$ the set of black points, and by $R=S\backslash Q$ the set of white points. We now define an adjacency relation to identify neighbors $s'$ of $s$, i.e. $s\sim s'$ if $||s-s'||_1 = 1$. $\sim$ is a symmetric relation on $S\times S$. Note that only either $s$ or $s'$ can be a member of $Q$ if $s\sim s'$, but not both at the same time. In a first step, we therefore have $\lambda_{1,r} = \lambda_{0,r}$ and $\gamma_{1, q} = \lambda_{0,q}$ for $r\in R$ and $q\in Q$. We now define specific prediction and update operators by

\begin{equation}
\label{quinpred}
P(\lambda_{1,r}) = \max_{q\colon q\sim r} \lambda_{1,q}
\end{equation}
and
\begin{equation}
\label{quinupdeq}
U(\gamma_{1,q}) = \max\{0, \max_{r\colon r\sim q}\gamma_{1,r} \}.
\end{equation}

In practice, when one considers classical cartesian pixel or voxel grids, these neighboring relations boil down to build minima and maxima among the 4- and 6-connected neighbors, respectively.

This scheme is known as Max-Lifting \cite{Heijmans_2000}, but the maximum can also be exchanged by a minimum. The use of Max- or Min-Lifting can be directly translated into the morphological operators dilation or erosion, respectively. Schemes that use this kind of prediction and update operators are therefore also called \emph{Morphological Wavelets}. 

A major advantage of lifting schemes as above is the fact that calculations can be performed inplace due to separation into even/odd or black/white. Consider first the 2D case, which is just the front slice of Figure~\ref{check_qc}. After the first lifting, the image only consists of the green quincunx points, so
a $45$~degree tilted version (or with diagonal instead of horizontal and vertical neighboring relations). If we now want to perform a prediction and an update step on the tilted version, we can use the exact same neighboring relations (horizontal and vertical neighbors) to apply the wavelet transform. This is even mathematically justified: The dilation matrix that transforms a quincunx sublattice into a Cartesian lattice in two dimensions is a similarity matrix. Unfortunately, in higher dimensions we fail to find such a transformation. A scheme which 
\begin{itemize}
	\item allows for a 2-channel design (split into even and odd, black and white), 
	\item generates a Cartesian lattice again after $d$ iterations, 
	\item \emph{and} is a similarity transform,
\end{itemize}

does not exist in any dimension larger than $d=2$ \cite{van2005multidimensional}. However, the third point is quite a crucial one: It guarantees that we can ``reuse'' neighboring relations we stated once. Lacking this similarity transform means that we need to borrow steps from another wavelet transform. But why does the scheme that worked so well in 2D fail in 3D? A rough intuition can be given by the Voronoi cells of the quincunx points.

Voronoi cells can be used to state an equivalent notion of neighborhood between points: Two points are neighbors if their Voronoi cells share exactly one face (one edge in 2D). Let us now consider the stepwise outcome of a wavelet transform on a quincunx grid. The first step, so splitting into black and white balls works fine in 2D and in 3D. We remain with two distinct sets from which we can predict and update. Performing prediction and update then yields an approximation on the positions of the black points. In 2D, we now just rotate the black points by $45$~degree, and once more we can use the previous neighboring relation, split the remaining points into two distinct sets, perform prediction and update steps on those and arrive eventually on a Cartesian grid with $\frac{1}{2}$ of image edge length in each coordinate direction, so an image whose total number of pixels is $\frac{1}{4}$ of the original. 

In 3D, the first step works similarly. But now considering the remaining points, we immediately unveil the issue in 3D. The Voronoi cells of the 3D quincunx grid have the shape of a rhombic dodecahedron, a twelve-faced polygon with rhombic faces, which means that every point has twelve neighbors. A similarity transform however requires the same amount of neighbors as in the Cartesian lattice, which are six. Even worse, the amount of twelve neighbors prevents us from dividing the remaining points into two distinct subsets, from which we can predict and update. We therefore borrow the idea of 5/3-LeGall filters~\cite{le1988sub}, which will not be introduced in detail here. 

The resulting Voronoi cell is now a cuboid, with two equal sides and the remaining one twice as long. The cell is not aligned with Cartesian coordinates. Clearly we loose isotropy at this stage. It may therefore be worth checking if the cuboidal cell should be aligned problem dependent to achieve better performance in whatever the wavelets are used for. This study will however not be part of this paper.

\subsection{Efficient Implementation of Morphological Wavelets}

In the previous section we saw that a generalization of splitting a one dimensional signal into even and odd parts is to split a multidimensional signal based on the quincunx grid. This can also be seen when we transform the multidimensional index to a linear one, i.e. a signal sampled at $n\times m$ or $n\times m\times l$ points, is transformed to a one dimensional signal by mapping
\begin{equation}
	(i,j)\mapsto i+j\cdot n,
\end{equation} 
or
\begin{equation}
    \label{multiind3d}
	(i,j,k) \mapsto i +(j + k\cdot n)\cdot m, 
\end{equation}
respectively. Then, the quincunx grid corresponds to the even and odd points of the transformed signals. However, transformation to 1D or modulo-checking for being even or odd are rather costly operations. We therefore extend the efficient implementation approach of~\cite{deZeeuw} to three dimensions. 

\begin{figure}
\hspace{1.5cm}
			\includegraphics[width=0.8\textwidth]{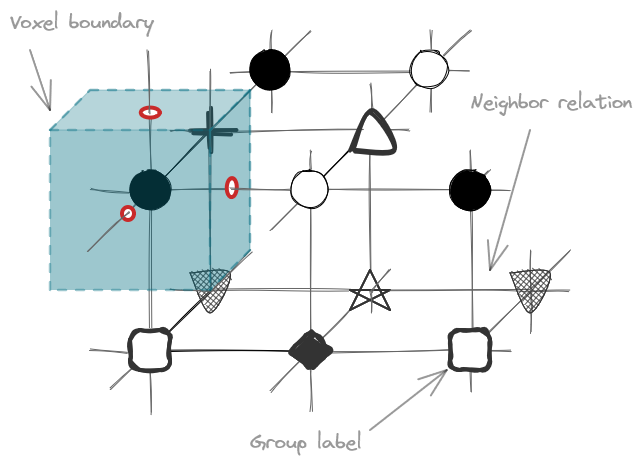}
	\caption[Voxel labels for efficient implementation]{Voxel labels for efficient implementation. To each voxel, a label is annotated. A voxel is depticted in dashed lines top left, solid lines represent neighboring relations between voxels. The presented pattern is repeated periodically in each coordinate direction over the whole image.}
	\label{Quincunxsymb}
\end{figure}

\begin{figure}
    \subfloat[Prediction for \inlg{ballw.png}]{
			\includegraphics[width=0.32\textwidth]{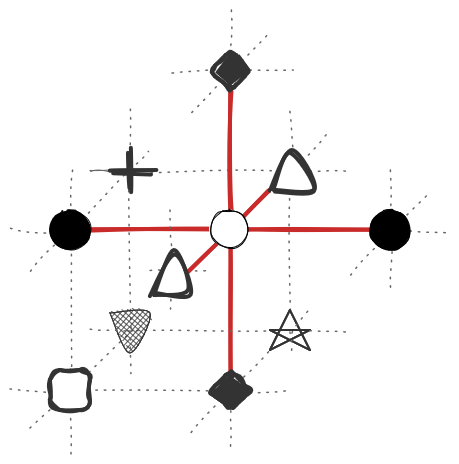}	
        \label{circ}
    }
    \hfill
    \subfloat[Prediction for \inlg{raute.png}]{
			\includegraphics[width=0.37\textwidth]{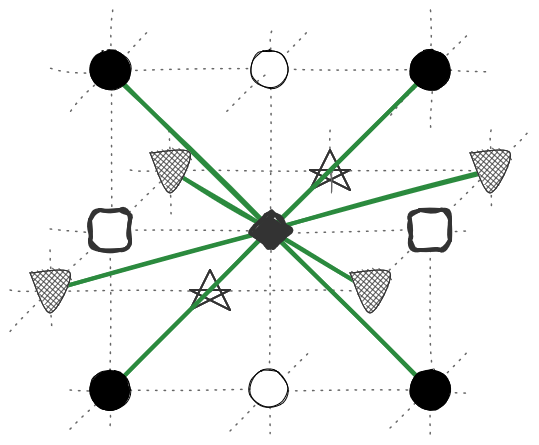}	
      \label{diamd1}
    }      
	\hfill
    \subfloat[Prediction for \inlg{cone.png}]{
			\includegraphics[width=0.2\textwidth]{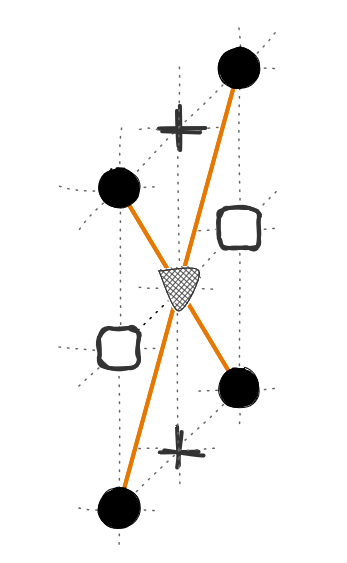}	
		\label{stard2}
	}
	
	\caption[Neighboring Relations for Morphological Wavelets]{Neighboring relations for morphological wavelets. Exemplary depicted for \inlg{ballw.png} in hv-lifting (Figure~\protect\subref{circ}), \inlg{raute.png} in d1-lifting (Figure~\protect\subref{diamd1}) and \inlg{cone.png} in d2-lifting (Figure~\protect\subref{stard2}). Actual neighboring relations are depicted in color, initial neighboring relations in dashed lines are kept for better orientation}
\end{figure}

For implementation, we do not split the lattice into two subsets, but eight. These are labelled by the symbols \inlg{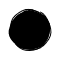}, \inlg{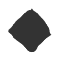}, \inlg{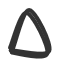}, \inlg{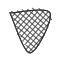}, \inlg{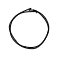}, \inlg{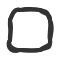}, \inlg{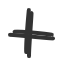}, \inlg{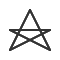} in Figure~\ref{Quincunxsymb}. Instead of modulo operations, this split can be efficiently performed by taking every second point in each dimension and only varying the starting points (either first or second point) in each dimension. We will call the application of the Morphological Wavelet Approximation to a Cartesian image grid \emph{Horizontal/Vertical Lifting} or \emph{hv}-Lifting. We start by splitting the image into $\big(\inlg{ballb.png}, \inlg{raute.png}, \inlg{traingle.png}, \inlg{cone.png}\big)$ and $\big(\inlg{ballw.png}, \inlg{cube.png}, \inlg{plus.png}, \inlg{star.png}\big)$. In this case, the neighboring relations are quite straight forward, as they are formed by the six neighbors in orthogonal coordinate directions. An example for the prediction of \inlg{ballw.png} can be found in Figure~\ref{circ}

Similarly, prediction steps are perfomed for $\big(\inlg{cube.png}, \inlg{plus.png}, \inlg{star.png}\big)$. The result is then used for an update step, again with straight-forward neighboring relations, for $\big(\inlg{ballb.png},\inlg{raute.png} , \inlg{traingle.png}, \inlg{cone.png}\big)$

We will call the transition from the dodecahedral grid to the cuboidal grid \emph{Diagonal Lifting 1} or \emph{d1-Lifting}. The points are now partitioned into $\big(\inlg{ballb.png}, \inlg{cone.png}\big)$ and $\big(\inlg{raute.png}, \inlg{traingle.png}\big)$. Again, we only depict the neighboring relation for prediction of \inlg{raute.png}. It can be found in Figure~\ref{diamd1}.

Application of the wavelet transform to the cuboidal grid is denoted by \emph{Diagonal Lifting~2} or \emph{d2-Lifting}. As only two sets are left the split is clear, the neighboring relations are however depicted in Figure~\ref{stard2}.

\subsection{Derivatives on the Lifted Grids}

In the previous subsection we learned that only in every third step we end up with a Cartesian grid in the wavelet transform. The other two steps produce grids that consist of cells in shapes of dodecahedra and tilted cuboids, respectively. However, if we want to perform motion estimation via OF on all scales, we need to estimate derivatives also on these special grids. We use the divergence theorem to do so -- a classical approach from finite element and volume methods. In what follows, we strongly follow the notation of \cite{Syrakos_2017} for unstructured grids, which clearly have our structured ones as special case. A cell of a grid will be denoted by $P$, with $\Omega_P$ its volume and $S_P$ its bounding surface, respectively. Then, for continuously differentiable $\Phi$ we have

\begin{equation*}
	\int_{\Omega_p} \nabla \Phi \dx = \int_{S_P}\Phi \vec{n}\,\mathrm d s. 
\end{equation*}

Our bounding surface $S_P$ consists of $F$ planar faces $S_f$, and each face has a constant unit normal, therefore
\begin{equation}
	\label{divtheorem}
	\int_{\Omega_p} \nabla \Phi \dx = \sum_{f=1}^F\left(n_f\int_{S_f}\Phi \,\mathrm d s\right). 
\end{equation}

Now apply the midpoint rule: It states that the mean value of a quantity over a cell $P$ (or face $f$) is equal to its value at centroid $\mathbf{P}$ of the cell (or $c_f$ of the face) plus a second order correction. We get

\begin{align*}
	\frac{1}{|\Omega_P|} \int_{\Omega_P} \nabla \Phi \dx &= \nabla \Phi(\mathbf{P}) + O(h^2) \\
	\frac{1}{|S_f|} \int_{|S_f|}  \Phi \,\mathrm d s &=  \Phi(c_f) + O(h^2),
\end{align*}

where $h$ is the characteristic grid spacing.
Together with the assumptions that multiplying by $\Omega_P$ causes an error of order $O(h^3)$ and $S_f$ of order $O(h^2)$, this yields

\begin{align*}
	\int_\Omega \nabla \Phi \dx &= \nabla \Phi(\mathbf{P})|\Omega_P| + O(h^5) \\
	\int_{S_f}  \Phi \,\mathrm d s &=  \Phi(c_f)|S_f| + O(h^4).
\end{align*}
Plugging this into (\ref{divtheorem}), we get
\begin{equation}
	\label{generalGradient}
	\nabla \Phi (\mathbf{P}) = \frac{1}{|\Omega|} \sum_{f=1}^{F} \Phi(c_f) |S_f| n_f + O(h^2).
\end{equation}

In the context of finite volume methods, one would now need to bother about evaluation of $\Phi$ at cell and face centers. But keep in mind that we calculate derivatives on a voxel grid, so $\Phi$ is just constant over the cell and the face.

Having now derived a method to compute gradients on arbitrary grids, we can apply the TV OF variant of Section~\ref{OF} to the quincunx grids. We just replace all differences by their corresponding components in the gradient, computed at every grid position as in equation \eqref{generalGradient}.

\subsection{MorphFlow}
\label{AlgoSec}

We will end this section with an exact description of the different steps of our algorithm. Due to the three-fold downsampling with morphological wavelets, the number of levels in the coarse-to-fine scheme has to be a multiple of $3$. Our convention is that start and endpoints of the morphological coarse-to-fine scheme are always classical cubic voxel settings. Each complete iteration step therefore starts with solving total variation regularized OF on the lowest scale. Next, the detail signals are retrieved, such that image information is available on a cuboidal voxel setting. To upscale the displacement fields from the coarser scale suitably, we act as if our fields stemmed from lossy compression with wavelets. That usually means that small detail coefficients in the wavelet approximation are set to $0$. The displacement fields from the previous scale are therefore retrieved with $0$ detail signal. Next, the procedure is repeated, but this time with image information on a dodecahedral voxel setting. If the final stage in the coarse-to-fine scheme is reached, an additional computation on the cubic voxel setting is performed. A pseudo-code description can be found in Algorithm~\ref{Algo}.

\begin{algorithm}[H]
	\caption{\label{Algo}MorphFlow} 
	\begin{algorithmic}[1]
	\Require  Images $I_0$, $I_1$, parameters $L_{\text{start}}$, $L_{\text{end}}$,$\tau$, $\lambda$, $\theta$, $i_{\text{warps}}$, $i_{\text{max}}$
	\Ensure Displacement $u$, $v$, $w$ 
        \State Construct wavelet decomposition $C$ of $L_{\text{start}}$ levels
		\For {$l=L_{\text{start}},L_{\text{start}}-3,\ldots L_{\text{end}}$}
		    \State Retrieve cubic image coefficients from C at level $l$
            \State Retrieve displacement field $(u,v,w)$ at level $l$ from $l-1$ with $0$-detail
			\For {$j=1,2,\ldots,i_{\text{warps}}$}
			    \State Compute $I_1(\ux+\ub^j)$, $\nabla I_1(\ux+\ub^j)$, $\rho(\ux+\ub^j)$,
			    \For {$k=1,2,\ldots,i_{\text{max}}$}
				    \State Calculate $\pb^{k+1}, \ub^{j, k+1}, \vb^{k+1}$ based on \eqref{finalupdatep}, \eqref{finapupdatestepu} and \eqref{finateupdatestepv}.
			    \EndFor
		    \EndFor
		\State Retrieve cuboidal image coefficients from C at level $l-1$
        \State Retrieve displacement field $(u,v,w)$ at level $l$ from $l-1$ with $0$-detail
	    \For {$j=1,2,\ldots,i_{\text{warps}}$}
			    \State Compute $I_1(\ux+\ub^j)$, $\nabla I_1(\ux+\ub^j)$, $\rho(\ux+\ub^j)$,
			    \For {$k=1,2,\ldots,i_{\text{max}}$}
				    \State Calculate $\pb^{k+1}, \ub^{j, k+1}, \vb^{k+1}$ based on \eqref{finalupdatep}, \eqref{finapupdatestepu} and \eqref{finateupdatestepv}.
                    \State Use \eqref{generalGradient} for gradient computation
			    \EndFor
		    \EndFor	
        \State Retrieve dodecahedral image coefficients from C at level $l-2$
        \State Retrieve displacement field $(u,v,w)$ at level $l$ from $l-1$ with $0$-detail
	    \For {$j=1,2,\ldots,i_{\text{warps}}$}
			    \State Compute $I_1(\ux+\ub^j)$, $\nabla I_1(\ux+\ub^j)$, $\rho(\ux+\ub^j)$,
			    \For {$k=1,2,\ldots,i_{\text{max}}$}
				    \State Calculate $\pb^{k+1}, \ub^{j, k+1}, \vb^{k+1}$ based on \eqref{finalupdatep}, \eqref{finapupdatestepu} and \eqref{finateupdatestepv}.
                    \State Use \eqref{generalGradient} for gradient computation
			    \EndFor
		    \EndFor	
  \EndFor
		
	\end{algorithmic} 
\end{algorithm}

\section{Datasets and Methodology}
\label{Datasets}
To demonstrate the performance of our algorithm, we use two experiments on concrete. The first one is a classical in situ test on refractory concrete. Due to the requirement of high refractoriness, the solid component consists of $70\%$ tabular alumina T60 aggregates, $20\%$  reactive alumina CTC50 and CT3000SG as fines, and only $5\%$ cement (alumina cement Secar 71). $6$ mass percent of water are added. The sample has a diameter of $24.5mm$ and a height of $25.2\,mm$. It has been extracted by core drilling from a larger sample of $50\,mm$ diameter and $50\,mm$ height. The voxel edge length of the CT images is $44\,\mu$m.  

Second, we use a three point bending test on concrete reinforced by a glass fiber rebar. The samples were produced at the chair of Concrete Structures and Structural Design at the department of Civil Engineering at RPTU Kaiserslautern-Landau. The concrete mixture used consists of CEM II 32.5 cement mixed with water in a water-cement ratio of $0.7$. Aggregates between 0 mm and 8 mm diameter were chosen. The in situ procedure has been performed at Fraunhofer EZRT in Fürth. The overall specimen size is $80\times80\times600$ mm, however a field-of-view scan was performed, such that the final image only covers $80\times80\times400 \, mm$. The final voxel edge length of the images is $128\,\mu$m. Note that in this work we show only excerpts of the sample to validate our method. A thorough mechanical evaluation based on our method will be postponed to future work.

The loading stages of this experiment were carefully chosen, which brings us in a rather unique position for in situ tests on brittle materials: We in fact can observe microcracking and early stage failure very close to the resolution limits. That means that we supposedly see cracks that have a width not more than a voxel or even lower. It is therefore an ideal test dataset to demonstrate the capacity of our algorithm to identify early stage failure. 

When applying the motion estimation algorithms, we  focus on two aspects: 
We show that for full scale computations our algorithm performs comparable to or even better than state-of-the-art methods. Furthermore, due to the use of wavelets, we can also use low scales to obtain a comparably cheap first analysis to -- for example -- decide with low computational effort, if a concrete sample in an in situ test is already fractured.

Therefore, our algorithm as presented in Section~\ref{AlgoSec} will be applied first to compute displacement fields at full resolution. To show that our approach indeed performs better than state-of-the-art methods from materials science, we also apply Augmented Lagrangian Digital Volume Correlation as comparison. 

Then we focus on the performance of morphological wavelets at low resolution scales. We show that neither 3DOF with Gaussian, nor 3DOF with Haar- or Daubechies-wavelet yield comparable results at similar resolution scales. Yet this performance is quite difficult to quantify. Our ''measures`` of choice will be \emph{residuals}, root-mean-square error (RMSE) and multilevel structural similarity (SSIM). Recall that motion estimation algorithms are supposed to reconstruct displacement vector fields~$\ub$ such that $I_0(\ux) = I_1(\ux+\ub)$. It is therefore very obvious to assess the performance of an algorithm based on this exact relation. Hence, one can consider the \emph{absolute value of the initial residual} \[r_0 = |I_0(\ux)-I_1(\ux)|,\] and the \emph{absolute value of residual} after warping,  \[r = |I_0(\ux)-I_1(\ux+\ub)|,\] where $I_1(\ux+\ub)$ was computed by interpolation at deformed coordinates. Note that we will often slightly carelessly call these quantities only \emph{initial residual} and \emph{residual}.

The RMSE is computed via

\begin{equation}
		\label{RMSEeq}
		\text{RMSE} \coloneqq \left(\frac{1}{N}\sum_x \left( I_1(\ux+\ub)-I_0(\ux)\right)^2\right)^{\frac{1}{2}}.
\end{equation}

Though the RMSE is widely used in image processing, it is well-known that it lacks of robustness towards outliers, see for example the discussion of Chai et al. \cite{chai2014root}. Yet, the main reason for Wang et al. to propose SSIM \cite{wang2002image} and later multi level SSIM \cite{wang2003multiscale} was that ''classical`` error measures like RMSE do not coincide well with human perception of similarity. 

To define SSIM, consider two discrete, non-negative signals $\mathbf{p} = \{p_i | i = 1, 2, \ldots, N\}$ and $\mathbf{q} = \{q_i | i = 1, 2, \ldots, N\}$  of length $N$ (for example image patches with consecutively numbered voxels as in \eqref{multiind3d}. The ingredients for SSIM are means $\mu_{\mathbf{p}}, \mu_{\mathbf{q}}$, variances $\sigma_{\mathbf{p}}^2$ and $\sigma_{\mathbf{q}}^2$, and covariance $\sigma_{\mathbf{p}\mathbf{q}}$. As mentioned, SSIM aims for mimicking the human perception, and it turns out, that $\mu_{\mathbf{p}}$ is an estimate of luminance, $\sigma_\mathbf{p}^2$ an estimate of contrast, and $\sigma_{\mathbf{p}\mathbf{q}}$ gives a tendency on how $\mathbf{p}$ and $\mathbf{q}$ vary together, so how structurally similar they are. Thus, the multiplicative components for SSIM are
\begin{align}
    l(\mathbf{p}, \mathbf{q}) &= \frac{2 \mu_{\mathbf{p}} \mu_{\mathbf{q}} + C_1}{\mu_{\mathbf{p}}^2 +\mu_{\mathbf{q}}^2+C_1} \\
     c(\mathbf{p}, \mathbf{q}) &= \frac{2 \sigma_{\mathbf{p}} \sigma_{\mathbf{q}} + C_2}{\sigma_{\mathbf{p}}^2 +\sigma_{\mathbf{q}}^2+C_2} \\
      s(\mathbf{p}, \mathbf{q}) &= \frac{\sigma_{\mathbf{p}\mathbf{q}} + C_3}{\sigma_{\mathbf{p}}\sigma_{\mathbf{q}}+C_3}, 
\end{align}
and 
\begin{equation}
    \label{SSIM}
    \text{SSIM}(\mathbf{p}, \mathbf{q})) = l(\mathbf{p}, \mathbf{q})^\alpha \cdot c(\mathbf{p}, \mathbf{q})^\beta \cdot s(\mathbf{p}, \mathbf{q})^\gamma.
\end{equation}
$C_1, C_2$ and $C_3$ are small constants to prevent numerical instabilities, and the exponents can be seen as weights, but are often just chosen such that $\alpha=\beta=\gamma = 1$. Clearly, the SSIM ranges between $0$ and $1$, and equals $1$, if both patches are the same. The SSIM index for a whole image is now the mean over all patches. It becomes multiscale, if the the SSIM is computed on different scales, i.e. on $M$ downscaled and filtered versions of the signals $\mathbf{p}$ and $ \mathbf{q}$

\begin{equation}
    \label{MSSIM}
    \text{ML-SSIM}(\mathbf{p}, \mathbf{q}) = l_M(\mathbf{p}, \mathbf{q})^{\alpha_M} \cdot \prod_{j=1}^Mc_j(\mathbf{p}, \mathbf{q})^{\beta_j} \cdot s_j(\mathbf{p}, \mathbf{q})^{\gamma_j}.
\end{equation}

To visualize the consequences of downscaling when considering cracks we use so called scanline plots. A scanline plot is a grey value profile along a horizontal or vertical line of one slice of the image. When applied at positions with local minima, it unveils very visibly how morphological wavelets preserve local minima best.

For both samples, we use strain as indicator for cracks. In our displacement fields, we expect discontinuities along the cracks. This means that strains in this area will be large. We use the following (mechanically slightly inaccurate) discrete approximation to local strain. Each component of the strain tensor $\varepsilon_{ij}$ is computed via

\begin{equation}
\label{straineq}
    \varepsilon_{ij} = \frac{1}{2}\left(\frac{\partial \ub_i}{\partial \ux_j }+ \frac{\partial \ub_j}{\partial \ux_i}\right).
\end{equation}
Thus, crack closure is a local maximum in the $(3,3)-$ component of the strain. As we have voxel-wise displacement fields, we can approximate this component easily by finite differences.

\section{Results and Discussion}
\label{Eval}

\begin{table}
	\centering
	\begin{tabular}{l | c c c c c c }
		\hline 
		Parameter & $L_{\text{start}}$ &  $\tau$ & $\lambda$ & $\theta$ & $i_{\text{warps}}$ & $i_{\text{max}}$ \\ \hline
        refractory & $12$ & $0.25$ & $25$ & $0.2$ & $20$ & $30$\\ 
        reinforced & $12$ & $0.25$ & $25$ & $0.2$ & $15$ & $30$\\\hline
        \end{tabular}
	\vspace{2mm}
	\caption{\label{paramTable}Parameters for MorphFlow used in the application examples.}
\end{table}

\begin{figure}
	\subfloat[Initial Residual]{                      
		\includegraphics[width=0.30\textwidth]{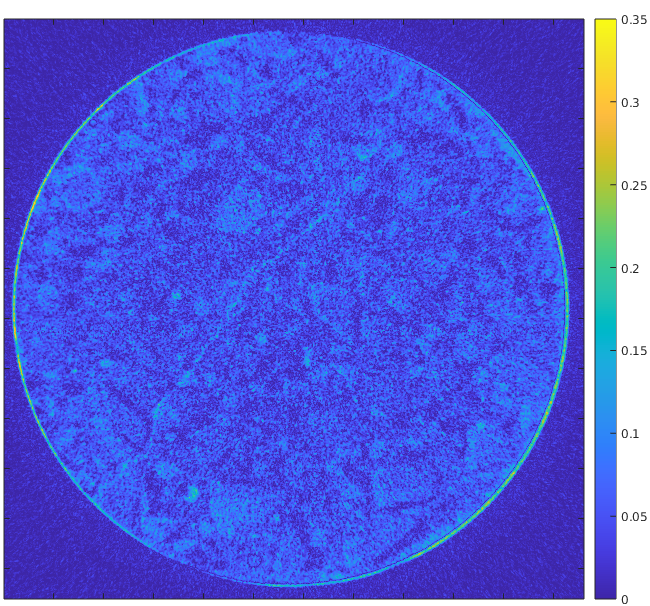}
		\label{res0KD2}
	}
	\hfill
	\subfloat[Residual MorphFlow]{
		\includegraphics[width=0.30\textwidth]{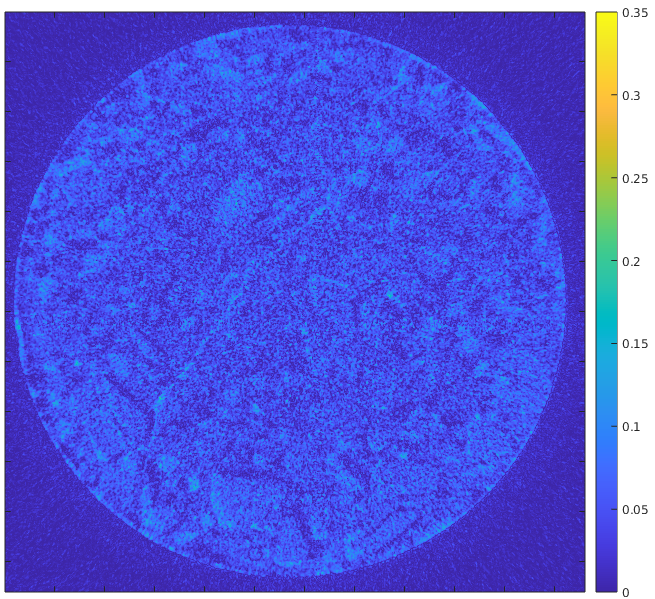}
		\label{res0KD4}
	}
	\hfill
	\subfloat[Residual ALDVC]{
		\includegraphics[width=0.30\textwidth]{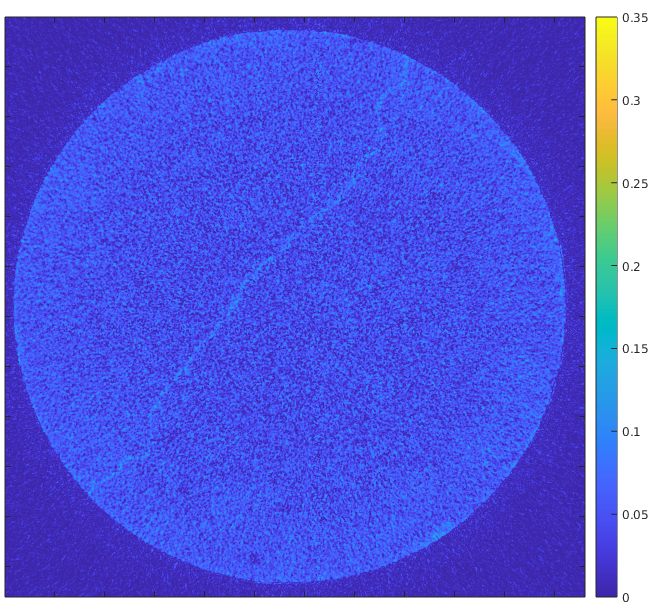}
		\label{res0KDmorph}
	}
	\caption{Sliceviews of the residuals of refractory concrete before and after warping at full resolution for MorphFlow and ALDVC.}
	\label{KDFresfull}
	\hspace{1cm}
\end{figure}

\begin{figure}
	\subfloat[$u$ MorphFlow]{                      
		\includegraphics[width=0.45\textwidth]{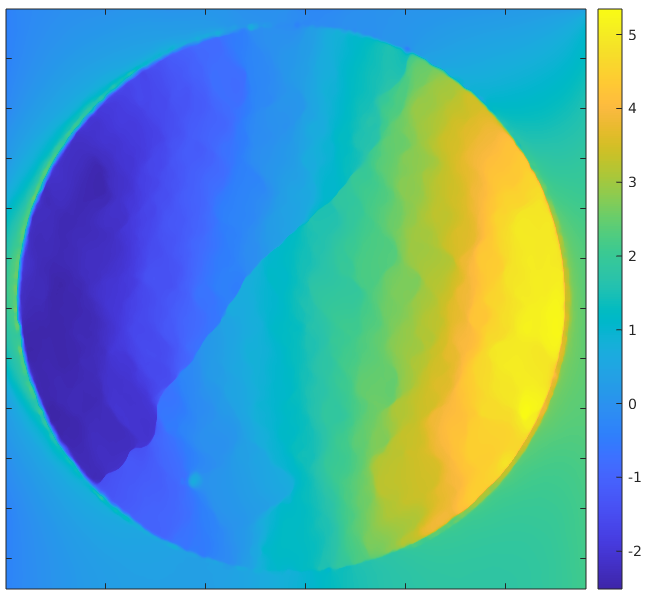}
		\label{KDFumorphfull}
	}
	\hfill
	\subfloat[$u$ ALDVC]{
		\includegraphics[width=0.45\textwidth]{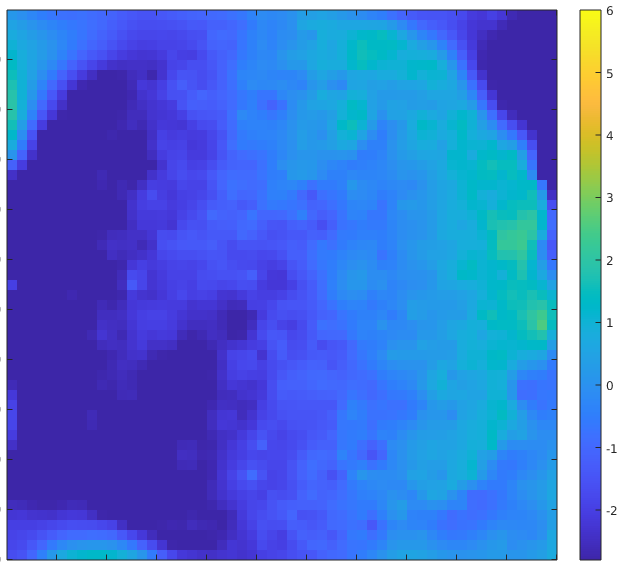}
		\label{KDFuALDVC}
	}
	\\
	\subfloat[$v$ MorphFlow]{
		\includegraphics[width=0.45\textwidth]{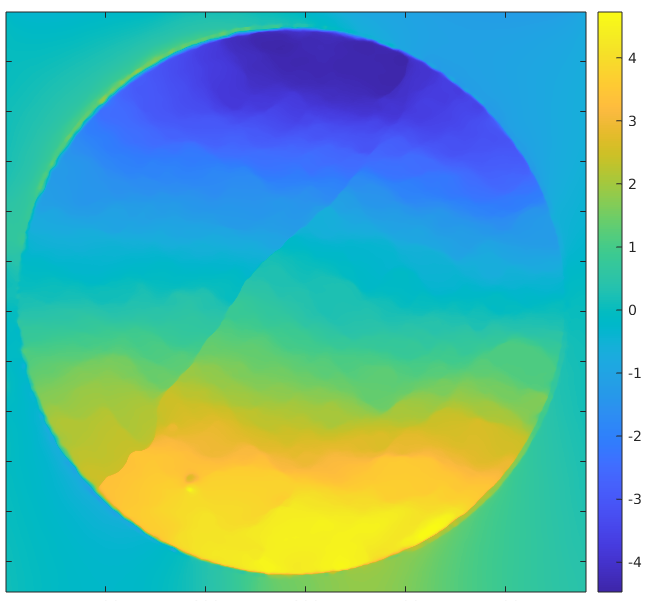}
		\label{KDFvmorphfull}
	}
	\hfill
    \subfloat[$v$ ALDVC]{
		\includegraphics[width=0.45\textwidth]{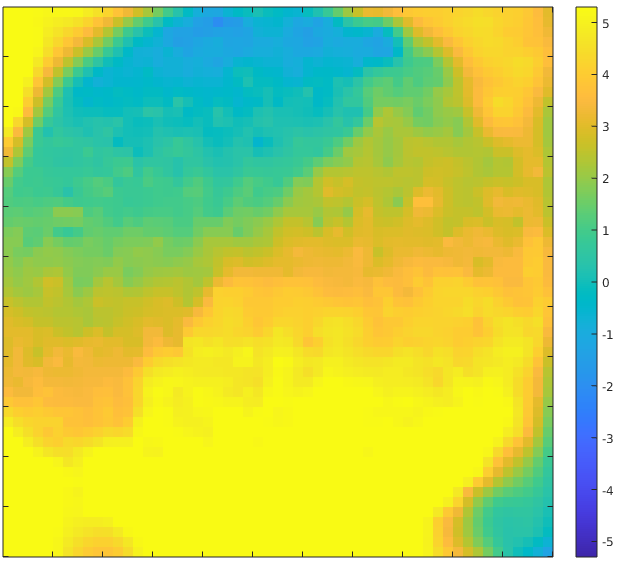}
		\label{KDFvALDVC}
	}
	\caption{Sliceviews of the displacement field components $u$ and $v$ of refractory concrete.}
	\label{KDFdispfull}
	\hspace{1cm}
\end{figure}

We want to start this section by confirming the findings of our previous work, namely that OF based motion estimations perform much better than state-of-the-art DVC based methods. To do so, we compare the calculated displacement of a full coarse-to-fine iteration of MorphFlow to, from our perspective, the best performing DVC method currently available, namely Augmented Lagrangian DVC (ALDVC) \cite{yang2020aldvc}. The parameters used for MorphFlow for the two datasets are given in Table~\ref{paramTable}. ALDVC was computed with a subvolume size of $40$ voxels and a subset size of $10$ voxels in each coordinate direction. Figure~\ref{KDFresfull} shows sliceviews of the full residuals. Visually, MorphFlow seems to perform slightly better, as the residual is in fact minimally darker on average. This is also verified by the RMSEs in Table~\ref{restable}. We can also observe the typical behavior of DVC-based methods, when examining Figure~\ref{KDFdispfull}. Though the crack is also visible up to some extent in the solution provided by ALDVC, the full-field displacement calculation produces much clearer and distinct fields, not only regarding the crack, but also regarding sample edges and boundary artifacts. They also appear to be a noisy, downsampled version of our displacement fields.

As already mentioned, our algorithm has very specific use cases in mind: It shall perform well when applied to low resolution approximations of concrete. In this evaluation we therefore compare our algorithm to several multiresolution schemes: the classical Gaussian pyramid, Haar wavelets and Daubechies Wavelets. Table \ref{restable} shows the RMSE for different multiresolution schemes, but at comparable stages. All values are computed for images that have roughly $1/4$ of the voxel edge length in each direction. Comparing the values immediately reveals another problem when comparing performances against each other: It lies in the nature of the wavelets that the further the approximation proceeds the larger the coefficients become. That is due to the assumption that large coefficients contribute significantly to the actual information content of the signal. Small coefficients on the other side are associated with noise. 

\begin{table*}
	\centering
	\begin{tabular}{l l | c c c | c c }
		\hline 
		\multicolumn{2}{c}{Data set} & \multicolumn{3}{c}{RMSE} & \multicolumn{2}{c}{ML-SSIM} \\
		& &  initial & after warp & decay & initial & after warp  \\ \hline
        {\multirow{2}{*}{Full resolution}} & MorphFlow & $0.0464$ & $0.040$ & $14\%$ & -- & --\\ & ALDVC & $0.0464$ & $0.0416$& $10\%$& -- & --\\ 
        \hline
        {\multirow{4}{*}{$1/4$ resolution}} & Haar & $0.3354$ & $0.2329$ & $23\%$ & $0.88$ & $0.90$ \\ & Daubechies & $0.3289$ & $0.2356$ & $29\%$ & $0.86$ & $0.91$\\ & Gauss & $0.0396$ & $0.0339$ & $14\%$ & $0.91$ & $0.88$ \\ & MorphFlow & $0.0479$ & $0.0408$ & $15\%$ & $0.89$ & $0.93$\\
	\end{tabular}
	\vspace{1mm}
	\caption{\label{restable} RMSE and ML-SSIM for refractory concrete at full resolution of the dataset and at downsampling of $1/4$ in each coordinate direction. For RMSE lower is better, for SSIM higher is better.}
\end{table*}

\begin{figure}
	\subfloat[Initial Haar]{                      
		\includegraphics[width=0.29\textwidth]{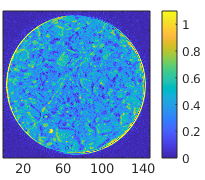}
		\label{res0db2}
	}
	\hfill
	\subfloat[Initial Daubechies]{
		\includegraphics[width=0.302\textwidth]{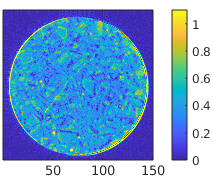}
		\label{res0db4}
	}
	\hfill
	\subfloat[Initial MorphFlow]{
		\includegraphics[width=0.302\textwidth]{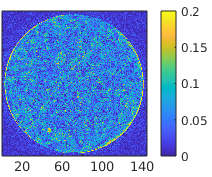}
		\label{res0morph}
	}
	\\
	\subfloat[After warping Haar]{
		\includegraphics[width=0.29\textwidth]{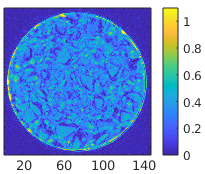}
		\label{res1db2}
	}
	\hfill
	\subfloat[After warping Daubechies]{
		\includegraphics[width=0.302\textwidth]{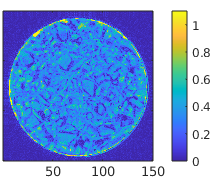}
		\label{res1db4}
	}
    \hfill
    \subfloat[After warping MorphFlow]{
		\includegraphics[width=0.302\textwidth]{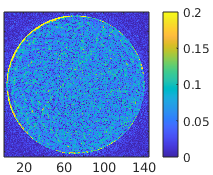}
		\label{res1morph}
	}
	\caption{Sliceviews of the residuals of refractory concrete before and after warping at reduced resolution. Haar, Daubechies and morphological wavelet approximations were applied twice to reduce the image edge length in each direction by a factor $4$. Note that the range of the color bar for Haar and Daubechies differs from MorphFlow due to the min- and max-preserving property of morphological wavelets. Ranges are set according to the $90$-percent quantile of all values in the initial residuals to guarantee similar impressions without weight on outliers.}
	\label{KDFres}
	\hspace{1cm}
\end{figure}

In contrast, morphological wavelets and Gaussian pyramids preserve the original image range. The RMSEs with respect to Haar and Daubechies wavelet approximations are therefore much higher. It is therefore more interesting to consider how much the RMSE actually \emph{decayed}. Therefore, the table also includes the initial RMSE for each dataset. 
Based on this, we can observe another fact: If we only look at the RMSE decay, our method performs better than Gaussian pyramids, but apparently not better than Haar or Daubechies wavelets. This is no surprise: As already mentioned, the RMSE is very susceptible to noise. That means, that the RMSE of a a pair of noisy images will be higher than the RSME of its denoised variants. Classical wavelets like Haar and Daubechies aim to suppress noise, as these are the highly oscillatory signal parts which are omitted first in signal decomposition. Morphological wavelets on the other side use operators that are likewise sensitive to noise, and even worse, may amplify it. This is also visually supported when looking at the residuals in Figure \ref{KDFres}, where \ref{res0db2}, \ref{res0db4}, \ref{res1db2} and  \ref{res1db4} look much smoother than \ref{res0morph} and \ref{res1morph}. 

It makes therefore sense to consider an error measure that is less susceptible to noise, the already mentioned ML-SSIM. Here in fact our method \emph{does} perform better -- the images are more similar with MorphFlow. Interestingly, the ML-SSIM regarding Gaussian smoothing even decays: When applied to concrete, classical coarse-to-fine methods destroy so much of the signal, that a motion estimation algorithm fails to recover any detail. 

\begin{figure}
	\subfloat[$u$ Haar]{                      
		\includegraphics[width=0.30\textwidth]{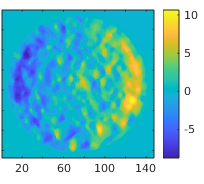}
		\label{KDFudb2}
	}
	\hfill
	\subfloat[$u$ Daubechies]{
		\includegraphics[width=0.317\textwidth]{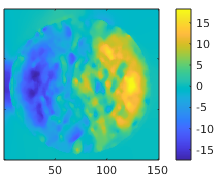}
		\label{KDFudb4}
	}
	\hfill
	\subfloat[$u$ MorphFlow]{
		\includegraphics[width=0.29\textwidth]{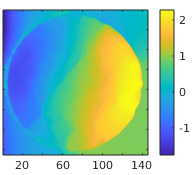}
		\label{KDFumorph}
	}
	\\
	\subfloat[$v$ Haar]{
		\includegraphics[width=0.30\textwidth]{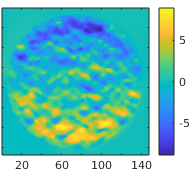}
		\label{KDFvdb2}
	}
	\hfill
	\subfloat[$v$ Daubechies]{
		\includegraphics[width=0.318\textwidth]{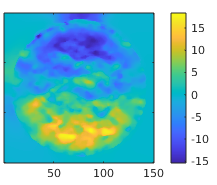}
		\label{KDFvdb4}
	}
    \hfill
    \subfloat[$v$ MorphFlow]{
		\includegraphics[width=0.29\textwidth]{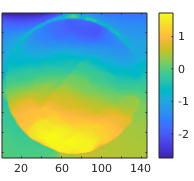}
		\label{KDFvmorph}
	}
	\caption{Sliceviews of the displacement field components $u$ and $v$ of refractory concrete at reduced resolution. Haar, Daubechies and morphological wavelet approximations were applied twice to reduce the number of voxels in each direction by a factor $4$. }
	\label{KDFdisp}
	\hspace{1cm}
\end{figure}

\begin{figure}
	\subfloat[$u$ Haar]{                      
		\includegraphics[width=0.40\textwidth]{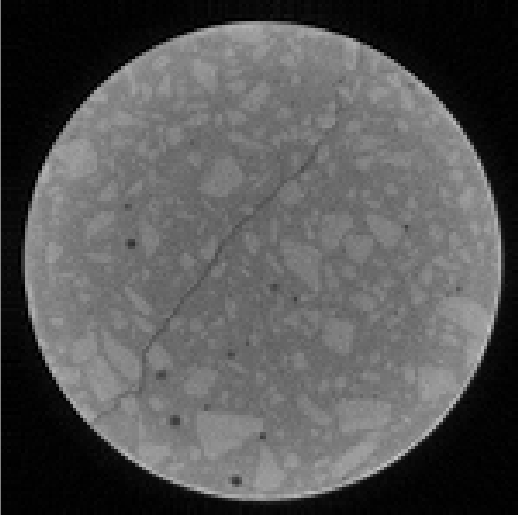}
		\label{dsdb4}
	}
	\hfill
	\subfloat[$u$ Daubechies]{
		\includegraphics[width=0.4\textwidth]{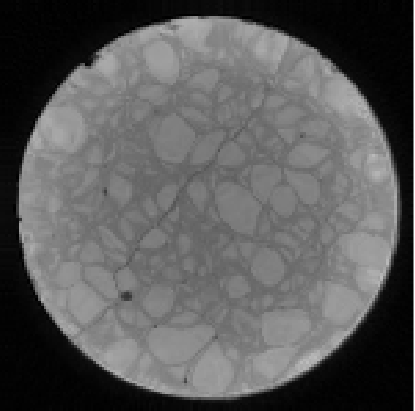}
		\label{dswdb4}
	}
	\caption{\label{blowup}Sliceviews of deformed image downscaled with Daubechies wavelet. Figure~\protect\subref{dsdb4} shows the deformed image with $1/4$ of the original length. Figure~\protect\subref{dswdb4} shows the same image, warped by the computed displacement fields. Note the ''blow  up`` of the grains. }
	\hspace{1cm}
\end{figure}

Residuals together with displacement fields now unveil another problem with classical wavelet approximations. Haar and Daubechies wavelets approximate the original image so well on the lower scales, that almost all details are still preserved. That unfortunately traps the algorithm in a local minimum, which can be seen by the wavy pattern in Figure~\ref{KDFdisp}. These give the correct idea, but produce unlikely valleys and hills in the displacement. Even further: If we compare the residuals from classical wavelet approaches with the one stemming from MorphFlow, we observe a bubble-like error pattern in the first two, whereas the latter just exhibits a rather regular noise pattern. If we now look exemplary at the deformed image based on the Daubechies wavelet in Figure~\ref{blowup}, we see that in fact the wavy pattern of the displacement field ''blows up`` the grains in the mortar phase. Clearly, this cannot represent a correct displacement field.

Morphological wavelets now unite the best of both worlds. They discard unwanted or unnecessary details such as grain boundaries, and dissolve them into noise. At the same time, they preserve the load induced detail of interest, namely the crack. Therefore, the displacement field appears to be very plausible, and nicely presents a jump, or a large variation along the crack. This will yield a maximum in strain -- exactly what is expected mechanically. 

\begin{figure}
 \subfloat[Original]{                      
		\includegraphics[width=1.0\textwidth]{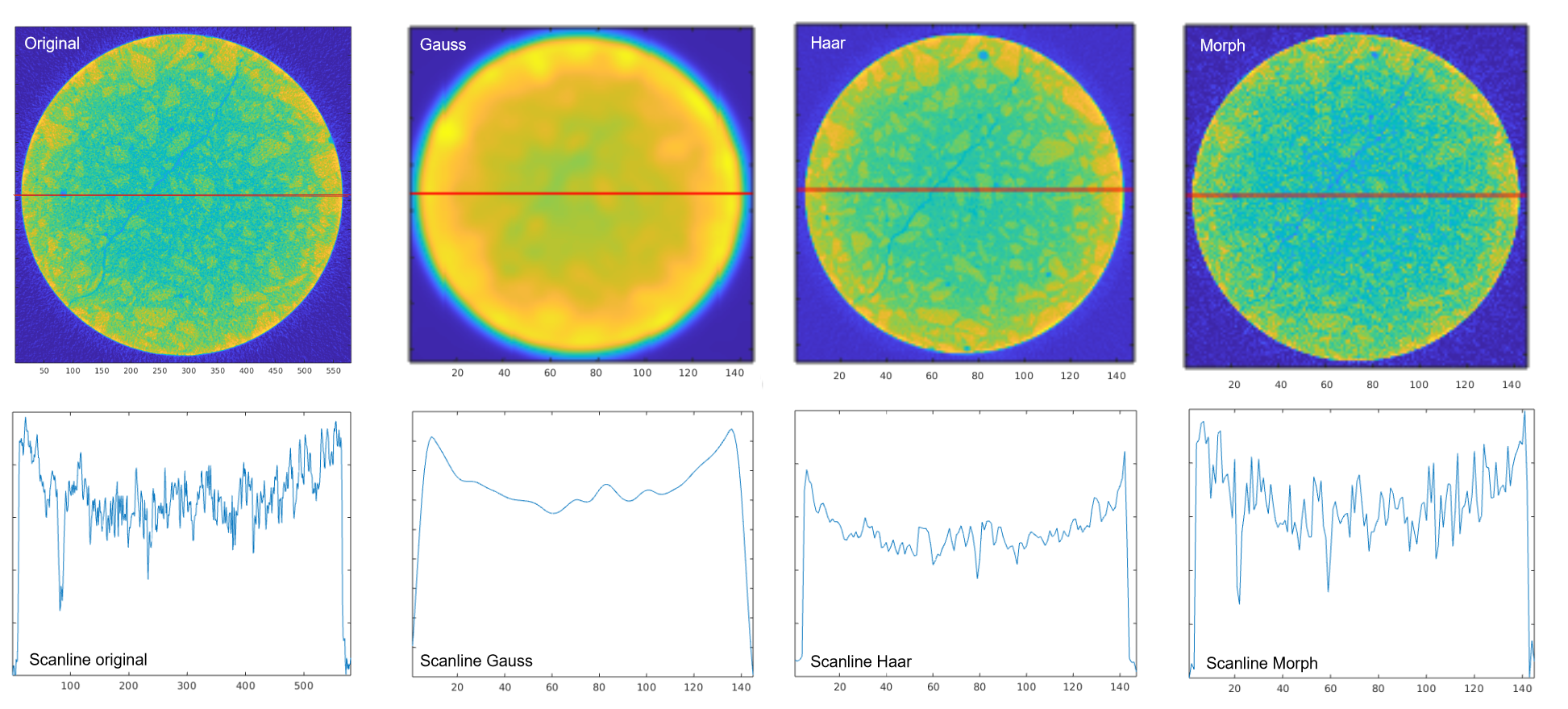}
		\label{KDForig_slimg}
	}
 
	\caption{Scanline representation of sliceviews of refractory concrete. Top row: Selected slices. The red line indicates the position for the grey value profile. Bottom row: Corresponding grey value profile, calculated along the red line.}
	\label{scanline}
	\hspace{1cm}
\end{figure}

Let us once more carefully observe Figure~\ref{KDFres}. Interestingly, the crack is not displayed as strongly in the residual as one would expect. Recalling, that several authors use bad residuals to indicate where the crack occurs \cite{lorenzoni_2020, mao_2019}, this is surprising at first sight. A different illustration can be found in Figure~\ref{scanline}. Here, we used the previously mentioned scanline representation, where the plot displays the grey value profile along the red line in the images. Clearly, Gaussian approximations smooth out all local minima and maxima. But surprisingly, this also holds true up to certain extent for classical wavelets. The local minimum in form of a crack around $x=60$ is damped quite significantly. This is also a reason, why motion estimation with classical wavelets does not produce plausible displacement fields.

\begin{figure}
 \subfloat[Rebar unloaded]{                      
		\includegraphics[width=0.41\textwidth]{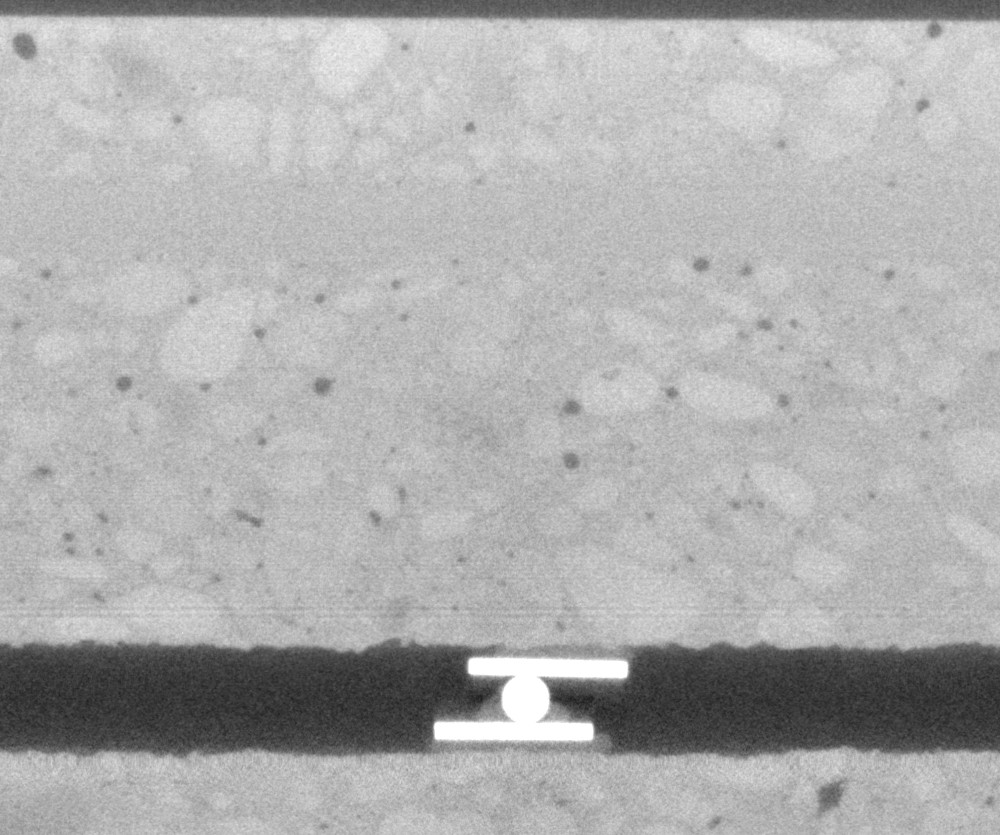}
		\label{rebarxun}
	}
    \hspace{0.75cm}
	\subfloat[Rebar loading step 1]{                      
		\includegraphics[width=0.41\textwidth]{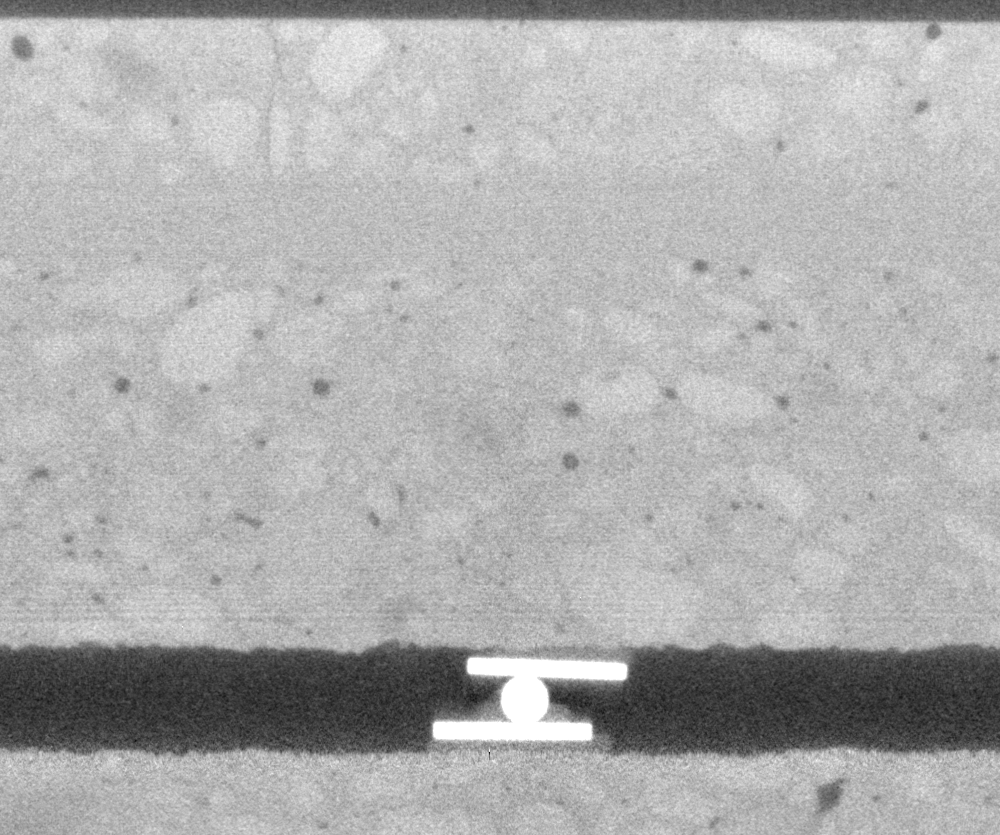}
		\label{rebarx1}
	}
    \\
    \subfloat[Rebar loading step 2]{
		\includegraphics[width=0.41\textwidth]{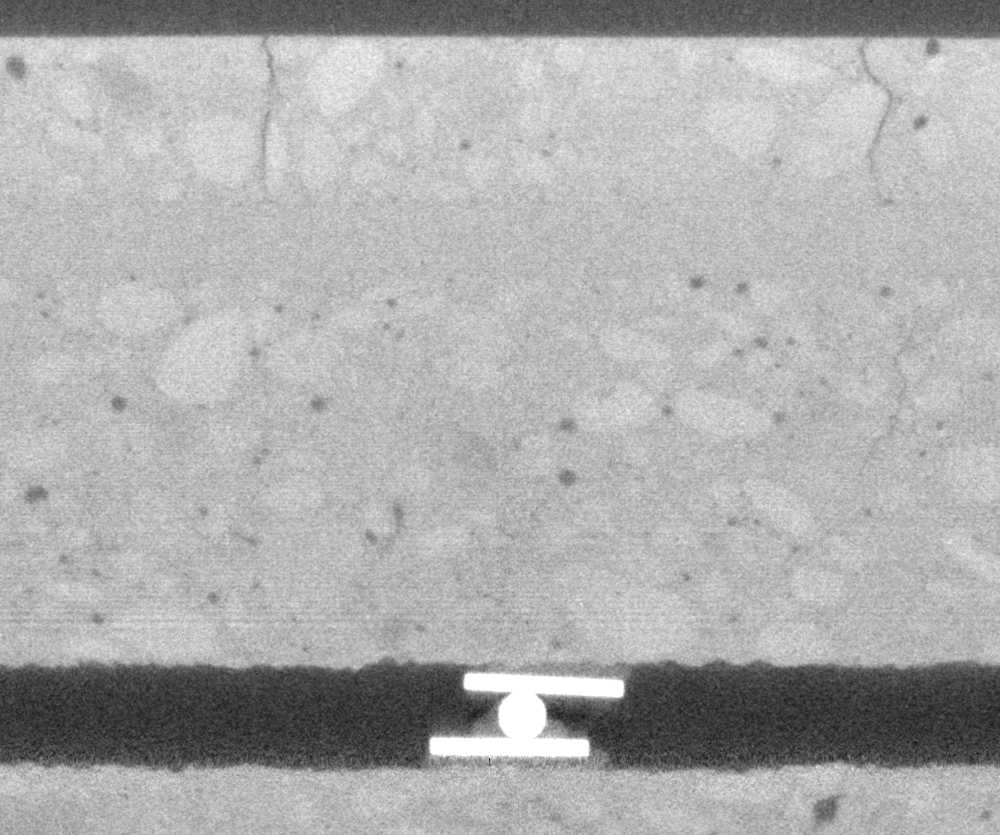}
		\label{rebarx2}
	}
	\hfill
	\subfloat[Displacement in w-direction (horizontally, (a) to (b))]{
		\includegraphics[width=0.47\textwidth]{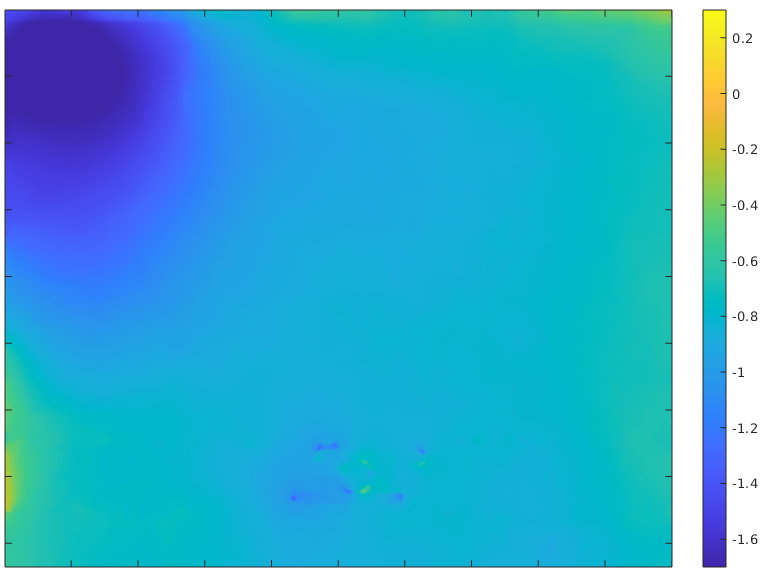}
		\label{rebardispx2}
	}
	
	\caption{Sliceviews of reinforced concrete. This sliceview allows to spot the glass fiber rebar in the upper third of the image. In both loading stages, the crack forms around the rebar. Figure~\protect\subref{rebardispx2} shows the displacement in w-direction (horizontal in this view) between unloaded and first loading stage.}
	\label{rebarsl}
	\hspace{1cm}
\end{figure} 

The second sample now shows how well our method predicts cracks already at low stages. Figure~\ref{rebarsl} shows a different sliceview with an additional loading stage. Again, the crack is difficult to spot in Figure~\ref{rebarx1}, but clearly visible in Figure~\ref{rebarx2}. The displacement field in w-direction in Figure~\ref{rebardispx2} also indicates a relatively small, smooth displacement. However, if we now calculate strain as described in \eqref{straineq}, we see a clear and distinct local maximum along the crack surface. Even further, up until the third downscaling step we still see a distinct local minimum at the crack location. This reduction of complexity is one of the core contributions of our algorithm: To reveice a decent first indication and approximation of the crack we require only  $0.2\%$ of the original number of voxels, as the third downscaling step reduces the number of voxels in each coordinate direction by a factor 8.

\begin{figure}
 \subfloat[Strain at full resolution]{                      
		\includegraphics[width=0.45\textwidth]{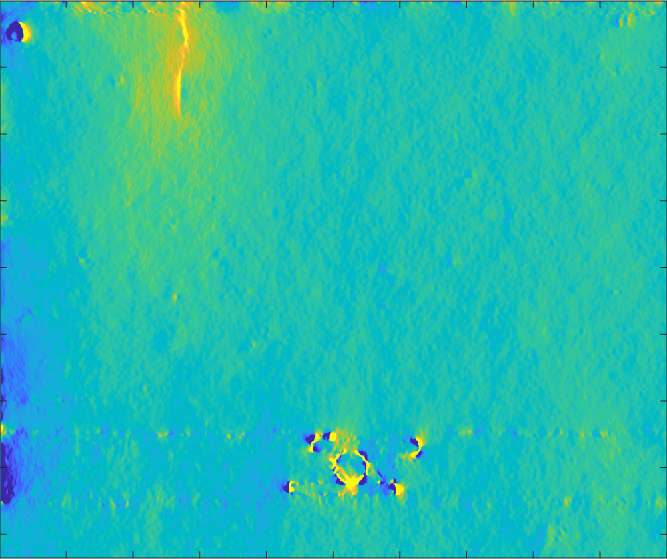}
	}
 \hspace{0.6cm}
 \subfloat[Strain at 1/2 resolution]{                      
		\includegraphics[width=0.455\textwidth]{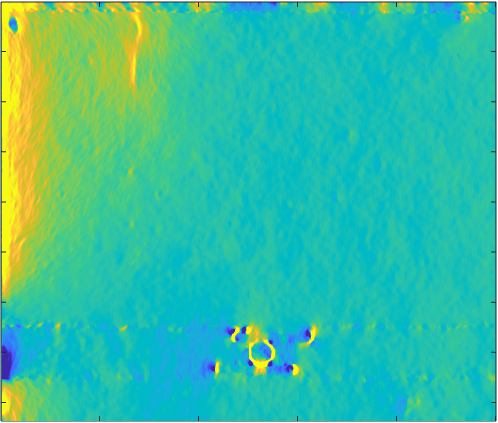}
	}
	\\
 \subfloat[Strain at 1/4 resolution]{                      
		\includegraphics[width=0.45\textwidth]{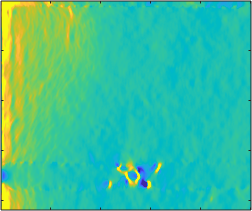}
	}
	\hfill
 \subfloat[Strain at 1/8 resolution]{                      
		\includegraphics[width=0.46\textwidth]{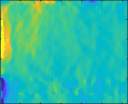}
	}
	\caption{Sliceviews of strains calculated from displacement between unloaded and first loading step. depicted at the same slice as in Figure~\ref{rebarsl}. The strain values show good agreement with literature (see for example \cite{shen2019tensile}. Yellow indicates strain maxima around $4\times 10^{-4}$, turquoise indicates zero strain, and blue negative strain. Note that strain minima are mainly found at clamping and boundaries and can therefore be neglected.}
	\label{rebarxstrain}
	\hspace{1cm}
\end{figure}

\begin{figure}
 \subfloat[Strain at full resolution]{                      
		\includegraphics[width=0.45\textwidth]{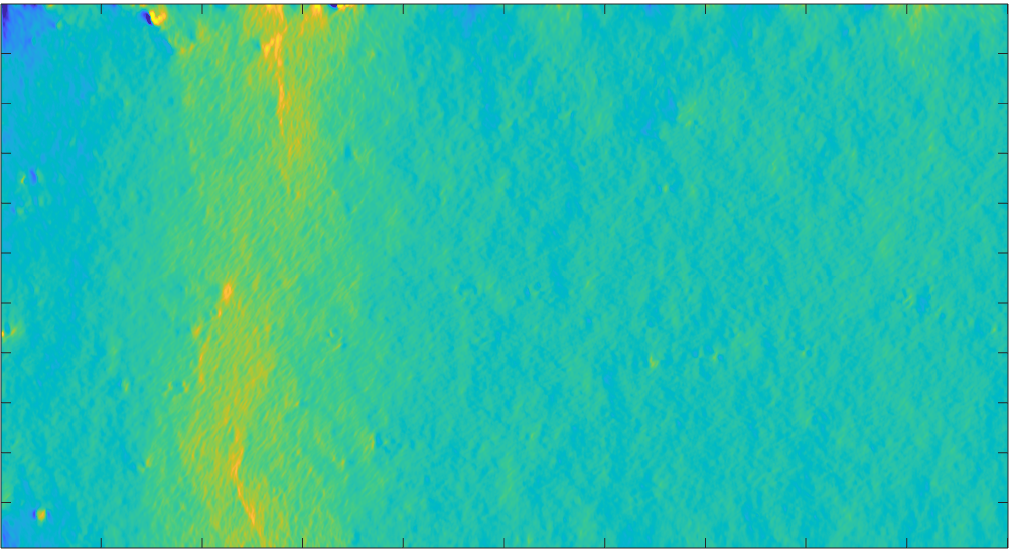}
	}
 \hspace{0.7cm}
 \subfloat[Strain at 1/2 resolution]{                      
		\includegraphics[width=0.455\textwidth]{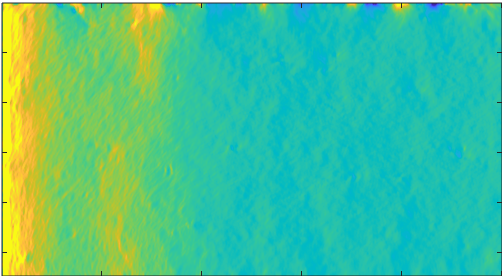}
	}
	\\
 \subfloat[Strain at 1/4 resolution]{                      
		\includegraphics[width=0.45\textwidth]{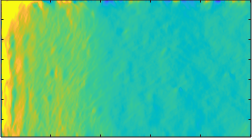}
	}
	\hfill
 \subfloat[Strain at 1/8 resolution]{                      
		\includegraphics[width=0.445\textwidth]{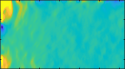}
	}
	\caption{Sliceviews of strains calculated from displacement between unloaded and first loading step depicted at the same slice as in Figures~\ref{gfk0sl} and \ref{gfk2sl}. The strain values are similar to Figure~\ref{rebarxstrain}. Though difficult to spot by the human eye, the strain clearly indicates a crack on all downscaling steps.}
	\label{rebarystrain}
	\hspace{1cm}
\end{figure}

\section{Conclusion}
\label{Conclusion}

We presented a novel approach to multiresolution motion estimation based on optical flow and morphological wavelets. We showed that our algorithm is extraordinary useful when applied to concrete datasets. Multiresolution schemes have been applied successfully in motion estimation numerous times. Most applications however involve real world camera images. In this area, Gaussian image pyramids are considered state-of-the-art. In applications such as in situ investigations of concrete they unfortunately miserably fail. Other ''classical`` multiresolution schemes like wavelets on the other side turned out to preserve too much detail to allow for accurate and plausible displacement fields. We showed that morphological wavelets unify the best of both approaches: They preserve the detail of interest -- the crack -- while removing unnecessary detail, such as aggregates.

We showed that with this approach very accurate displacement fields can be achieved. Even further, due to there signal approximating nature, using morphological wavelets can be considered as calculations on compressed data. Therefore, we also investigated how good low resolution approximations perform in comparison. Morphological wavelets turned out to be superior once more. They allow for crack indicators on a fraction of the original number of voxels and even performed well for cracks with a width less than a voxel.

\section*{Acknowledgements}
Funding by the Research Initiative of the Federal State of Rhineland-Palatinate through the priority area Mathematics Applied to Real-World Problems (MathApp) is gratefully acknowledged.
This work was supported by the German Federal Ministry of Education and Research (BMBF) [grant number 05M2020 (DAnoBi)].
We thank Jana Hub\'alkov\'a from TU Bergakademie Freiberg for providing the refractorary concrete samples and image data. 
We thank Szymon Grzesiak from RPTU Kaiserslautern-Landau for providing the samples for the experiments on the reinforced concrete and Michael Salamon from Fraunhofer EZRT for scanning.

\clearpage
\bibliography{bib}

\end{document}